\documentclass[
11pt, % Main document font size
a4paper, % Paper type, use 'letterpaper' for US Letter paper
oneside, % One page layout (no page indentation)
headinclude,footinclude, % Extra spacing for the header and footer
]{scrartcl}

\usepackage[
nochapters, % Turn off chapters since this is an article        
beramono, % Use the Bera Mono font for monospaced text (\texttt)
eulermath,% Use the Euler font for mathematics
pdfspacing, % Makes use of pdftex’ letter spacing capabilities via the microtype package
dottedtoc % Dotted lines leading to the page numbers in the table of contents
]{classicthesis} % The layout is based on the Classic Thesis style

\usepackage{arsclassica} % Modifies the Classic Thesis package
\usepackage[T1]{fontenc} % Use 8-bit encoding that has 256 glyphs
\usepackage[utf8]{inputenc} % Required for including letters with accents
\usepackage{tgbonum}
\usepackage{graphicx} % Required for including images
\graphicspath{{../}} % Set the default folder for images
\usepackage{enumitem} % Required for manipulating the whitespace between and within lists
\usepackage{lipsum} % Used for inserting dummy 'Lorem ipsum' text into the template
\usepackage{amsmath,amssymb,amsthm} % For including math equations, theorems, symbols, etc
\usepackage{varioref} % More descriptive referencing

\usepackage{url}            % simple URL typesetting
\usepackage{amsfonts}       % blackboard math symbols
\usepackage{nicefrac}       % compact symbols for 1/2, etc.
\usepackage{microtype}      % microtypography
\usepackage[round]{natbib}
\usepackage{graphicx}
\usepackage{import}
\usepackage{lineno}
\usepackage{easyReview} % for \comment, \alert, etc.s
\usepackage{setspace}
\usepackage{geometry}
\hypersetup{% hyperrefオプションリスト
	setpagesize=false,
	bookmarksnumbered=false,%
	bookmarksopen=false,%
	colorlinks=false,%
}

\theoremstyle{definition} % Define theorem styles here based on the definition style (used for definitions and examples)

\theoremstyle{plain} % Define theorem styles here based on the plain style (used for theorems, lemmas, propositions)

\theoremstyle{remark} % Define theorem styles here based on the remark style (used for remarks and notes)

\title{\normalfont{A comparison of an operational wave-ice model product and drifting wave buoy observation in the central Arctic Ocean: investigating the effect of sea ice forcing in thin ice cover}} % The article title
\author{Takehiko Nose\textsuperscript{1}*, 
	Jean Rabault\textsuperscript{2},
	Takuji Waseda\textsuperscript{1},
	Tsubasa Kodaira\textsuperscript{1},\\
	Yasushi Fujiwara\textsuperscript{1,3},
	Tomotaka Katsuno\textsuperscript{1},
	Naoya Kanna\textsuperscript{4},\\
	Kazutaka Tateyama\textsuperscript{5},
	Joey Voermans\textsuperscript{6},
	\& Tatiana Aleekseva\textsuperscript{7,8}.
} % The article author(s) - author affiliations need to be specified in the AUTHOR AFFILIATIONS block

\date{} % An optional date to appear under the author(s)

\begin{document}
\begin{NoHyper}	
%\rmfamily	
%----------------------------------------------------------------------------------------
%	HEADERS
%----------------------------------------------------------------------------------------

\renewcommand{\sectionmark}[1]{\markright{\spacedlowsmallcaps{#1}}} % The header for all pages (oneside) or for even pages (twoside)
\lehead{\mbox{\llap{\small\thepage\kern1em\color{halfgray} \vline}\color{halfgray}\hspace{0.5em}\rightmark\hfil}} % The header style

\pagestyle{scrheadings} % Enable the headers specified in this block

%----------------------------------------------------------------------------------------
%	TABLE OF CONTENTS & LISTS OF FIGURES AND TABLES
%----------------------------------------------------------------------------------------

\maketitle % Print the title/author/date block

%----------------------------------------------------------------------------------------
%	AUTHOR AFFILIATIONS
%----------------------------------------------------------------------------------------

\let\thefootnote\relax\footnotetext{\textsuperscript{1}Department of Ocean Technology, Policy and Environment, Graduate School of Frontier Sciences, The University of Tokyo, Kashiwa Chiba Japan
}
\let\thefootnote\relax\footnotetext{\textsuperscript{2} 
Norwegian Meteorological Institute, IT Department, P.O. Box 43, Blindern N – 0313 OSLO
}
\let\thefootnote\relax\footnotetext{\textsuperscript{3} 
	Now at Kobe University
}
\let\thefootnote\relax\footnotetext{\textsuperscript{4} 
	The Atmosphere and Ocean, Research Institute, The University of Tokyo, Kashiwa Chiba Japan
}
\let\thefootnote\relax\footnotetext{\textsuperscript{5} 
	Kitami Institute of Technology, Kitami Hokkaido Japan
}
\let\thefootnote\relax\footnotetext{\textsuperscript{6} 
	The University of Melbourne, Melbourne Australia
}
\let\thefootnote\relax\footnotetext{\textsuperscript{7} 
	Arctic and Antarctic Research Institute, Saint Petersburg Russia
}
\let\thefootnote\relax\footnotetext{\textsuperscript{8} 
	 Space Research Institute, Moscow, Russia
}

\let\thefootnote\relax\footnotetext{*\texttt{tak.nose@edu.k.u-tokyo.ac.jp} }

%----------------------------------------------------------------------------------------
%	ABSTRACT
%----------------------------------------------------------------------------------------

\section*{Abstract} % This section will not appear in the table of contents due to the star (\section*)
Two drifting wave buoys were deployed in the central Arctic Ocean, north of the Laptev Sea, where there are historically no wave observations available. 
An experimental wave buoy was deployed alongside a commercial buoy. The inter-buoy comparison showed that the experimental buoy measured wave heights and periods accurately; so the buoy data were used to study the predictability of a wave-ice model in this region.
The first event we focused was when both buoys observed a sudden decrease in significant wave heights $H_{m0}$. The sudden decrease was caused by the change of wind directions from along the ice edge to off-ice wind. This event was compared with the ARC MFC wave-ice model product, which underestimated the $H_{m0}$. The inaccurate model representation of an ice tongue located upwind of the buoys was found to constrain the available fetch for wave growth.
The second case was when the buoys entered ice cover as new ice formed; an on-ice wave event was observed and the buoy downwind measured 1.25 m $H_{m0}$.
In this instance, the ARC MFC wave-ice model product largely underestimated the downwind buoy $H_{m0}$. To investigate this event, model sea ice conditions were examined by comparing the ARC MFC sea ice forcing with the neXtSIM sea ice model product.
Our analysis revealed that the thin ice thickness distribution for ice types like young and grey ice, typically less than 30 cm, were not resolved. The ARC MFC model's wave dissipation rate has a sea ice thickness dependence, and therefore, it overestimates wave dissipation in thin ice cover; sea ice forcing that can resolve the thin thickness distribution is needed to improve the predictability.
This study provides an observational insight into better predictions of waves in marginal ice zones when new ice forms. 
%----------------------------------------------------------------------------------------

\vspace{5mm}
% keywords can be removed
\textbf{\textit{Keywords}} {drifting wave-ice buoy, ARC MFC wave-ice model, neXtSIM sea ice model, wave-ice interaction, ice thickness, and Arctic Ocean}

\section*{RUNNING HEAD}
Investigating the effect of sea ice forcing for a wave-ice model in thin ice cover

\section*{ABBREVIATIONS}
AMSR2 - Advanced Microwave Scanning Radiometer 2\\
ARC MFC - Arctic Monitoring and Forecast Center\\
CICE - Community Ice CodE\\
CMEMS - Copernicus Marin Environment Monitoring Service\\
HYCOM - Hybrid Coordinate Ocean Model\\
IMU - inertial measurement unit\\
MIZ - marginal ice zone\\
OMB - OpenMetBuoy\\
SIC- sea ice concentration\\
SIT - sea ice thickness\\
WAM - WAve Model\\

\newpage % Start the article content on the second page, remove this if you have a longer abstract that goes onto the second page

%----------------------------------------------------------------------------------------
%	INTRODUCTION
%----------------------------------------------------------------------------------------

\doublespacing
\nolinenumbers

\section{Introduction}\label{intro}
Wave-ice interaction research is attracting renewed attention due to declining Arctic Ocean sea ice cover and a conjecture that waves in ice have an influence on the climate system \citep{Squire2018,Squire2020}.  Despite recent advancement in model parameterisation since one of the most intensive waves-in-ice data collection campaign of \cite{Thomson2018}, incorporating the effects of waves in sea ice models and vice versa remains a challenging problem.
In MIZs where sea ice field is heterogeneous and waves most dynamic, the problem becomes even more challenging as accurate sea ice condition is difficult to obtain. Uncertainty arising from SIC estimates (from passive microwave radiometers) used as sea ice forcing for wave-ice models can even overwhelm the uncertainty of wave-ice interaction parameterisations \citep{Nose2020}. 
Due to these challenges, it can be argued that more observations are needed to better understand wave-ice interaction physics and improve the predictability of ocean waves in MIZs.

Utilising the recent advances of inexpensive electronics and their open source philosophy, \cite{Rabault2022} developed and introduced a low-cost, easy to build wave-ice buoy called OMB.
The OMB applies a 6 degrees of freedom IMU to measure vertical ocean surface motion. The vertical surface oscillation can then be used to obtain power spectral density, from which integrated wave statistics like the wave height and periods can be calculated. In September 2021, our research team joined the NABOS campaign (\url{https://uaf-iarc.org/nabos-cruises/}) onboard R/V Akademik Tryoshnikov, and we deployed a prototype of \citet{Rabault2022}'s OMB  alongside a commercial wave buoy in the central Arctic Ocean where there are historically no wave observations available.
In this study, we evaluated our prototype OMB with a commercial wave buoy; we then used the buoy dataset to study the wave-ice model predictability in this region.

To simulate the ice effects on waves, we first need to consider the length scale between $\lambda $ and $D$ \citep{Linton2010} where $D$ could be considered the diameter of an ice floes and $\lambda$ is the wavelength.
For an ice sheet and grease ice where $\lambda < D$ and $\lambda >> $ SIT, the sea ice can be modelled as a thin elastic ice layer where waves propagate under ice, and the ocean-ice interface is where the dissipation occurs. When $\lambda \approx D$, wave attenuation is understood to be dominated by a conservative process known as the scattering mechanism, which was first observed by \cite{Wadhams1975}. 
When $\lambda > D$, waves propagate through smaller sea ice floes and dissipation occurs in many forms \citep{Squire2018}; this type of ice cover can be modelled as a semi-infinite viscous ice layer, in which effective material properties can be tuned to reproduce the aggregate effect of various sea ice effects on waves \citep{Squire2018}.

The $\lambda > D$ regime is primarily observed in the MIZ. This is the length scale of interest to our wave buoy observation because the ice charts indicated that the wave buoys were located near and in new and young ice, with typical thickness less than 30 cm, during the observation period (see \ref{icechart}). 
For the purpose of a model-observation comparison, we used an operational model product named ARC MFC\textsuperscript{+}\footnote{\textsuperscript{+}CMEMS product name: ARCTIC\_ANALYSIS\_FORECAST\_WAV\_002\_014. Data downloaded on 18 Jan 2022.}.

Because the ARC MFC model's wave dissipation parameterisation \citep{Sutherland2019} is suited for modelling waves in the MIZ thin ice cover, we expect reasonable model agreement with the observation. Our objective in this study is to elucidate the predictability of the ARC MFC wave-ice model in the central Arctic Ocean by applying the drifting wave buoy observations and other model products ECMWF HRES (wave) and neXtSIM (sea ice). Specifically, we focus on how sea ice edge and SIT representations in the sea ice forcing affect wave predictions when new ice is forming.

We describe the buoy observation details including sensor types and platforms as well as the inter-buoy comparison analysis in Section~\ref{buoys}. This is followed by a description of the numerical models used in this study in Section~\ref{model}. In Section \ref{iceeffectsonwaves}, we carry out the model-observation comparison using the ARC MFC wave-ice model. In section \ref{discussion}, we discuss the Section \ref{iceeffectsonwaves} results by exploring how different sea ice forcing can affect wave-ice models. Conclusions follow in Section \ref{conclusion}.

\section{Drifting wave buoy observation} \label{buoys}
\subsection{Wave buoy sensor and platform description}
During the 2021 NABOS campaign, we deployed the first prototype of \cite{Rabault2022} wave-ice buoy (herein referred to as Zeni-v2021) alongside a SOFAR Spotter (herein referred to as SPOT-1386). 
Both Zeni-v2021 and SPOT-1386 buoys measure ocean surface motion, but using different technology.

Zeni-v2021 was recently introduced in \citet{Rabault2022}.
The OMB electronic components for detecting ocean surface motion is a 6 degrees of freedom accelerometer + gyroscope IMU, ST ISM330DHCX (\url{https://www.st.com/en/mems-and-sensors/ism330dhcx.html}). The attitude heading reference system correction via sensor fusion of three-axis accelerations and angular rates produces true vertical acceleration \citep{Rabault2022}, which is integrated twice to estimate the surface elevation. 

Despite the name being a "buoy", the OMB is a sensor unit that is housed in a waterproof enclosure (see Fig. A1 of \cite{Rabault2022} for example), i.e., it is not designed as a floating platform in water. Therefore, the primary deployment method of OMBs is that the sensor unit be placed on an ice floe, and the ice floe becomes the sensor's floating platform (thereby the terminology "wave-ice buoy"). At the time of the buoy preparation (July 2021), however, the sea ice extent was following the 2012 record low sea ice extent\textsuperscript{+}\footnote{\textsuperscript{+}although by August, the sea ice extent reduction plateaued, resulting in a field observation conundrum that there were limited opportunities for open water deployment. For a quick reference of the sea ice extent, please visit \url{https://ads.nipr.ac.jp/vishop/\#/extent} and plot 2012 and 2021 sea ice extents.}. Accordingly, an alternative deployment method was devised by housing the sensor in a floating platform.
The ad hoc platform was a Zenilite GPS tracker enclosure (\url{https://www.zenilite.co.jp/english/prod/new-chikuden.html}), which was conveniently designed to house the OMB electronic components (see Appendix B of \citet{Rabault2022}). The Zenilite GPS drifting buoy has a diameter of 340 mm, and is 300 mm in height and weighs \textasciitilde6 kg. 

SPOT-1386 is a commercially sold drifting wave buoy; it is a proven technology with thousands of them currently deployed in the world oceans. The dimensions are similar to the Zenilite GPS drifter, which are 420 mm width by 310 mm in height and weighs 5.3 kg (7.4 kg with a ballast); therefore, SPOT-1386 is an appropriate benchmark for Zeni-v2021. Spotters are based on a proprietary firmware that uses a GPS/GNSS receiver to get the device horizontal and vertical displacements.

\subsection{Buoy deployment}
The wave buoys were deployed adjacent to an ice edge in the central Arctic Ocean north of the Laptev Sea (81.915$^\circ$ N, 118.763$^\circ$ E) at around UTC 05:05 15 September 2021. The deployed location is shown in Fig. \ref{fig1}, and the sea ice conditions as observed onboard R/V Akademik Tryoshnikov on the day of the buoy deployment are shown in Fig.~\ref{figA1}.
SPOT-1386 battery life at high latitudes without solar charge is approximately 10 days; as such, the co-located deployment duration only lasted between 15 and 29 September. 
The buoy tracks for this period are shown in Inset A of Fig. \ref{fig1}, which is overlaid with the AMSR2  SIC \citep{ADS-AMSR2} contours.

The primary motive for the buoys being deployed at the same location was to validate Zeni-v2021 against SPOT-1386, and we anticipated that the buoys measure analogous wave signal for at least several days. For example, \citet{Waseda2018,Nose2018} describe the trajectories and wave statistics of two buoys deployed at the same location in the ice-free Beaufort Sea in 2016; the buoys drifted along similar tracks for \textasciitilde13 days when they measured analogous wave statistics.
For our observation, however, Zeni-v2021 and SPOT-1386 wave heights began deviating slightly merely 12 hours after the buoys were deployed. After 2 days of being deployed, wave heights and periods varied considerably, which indicates that the measured waves' evolution did not occur entirely over open ocean, i.e., sea ice affected how the waves evolved. 

\subsection{Co-located wave buoy measurements in thin ice field}\label{colocatedbuoy}
\textbf{Overview of the wave data:}
Figure \ref{fig2} presents an overview of the co-located buoy observation: the buoy distances, and wind, wave, and SIC conditions. Here, wave statistics derived from the vertical surface elevation are significant wave height $H_{m0}= 4\sqrt{m_0}$ where $ m_0 = \int_{f0}^{f1} S(f)df $ and wave periods (peak period $T_p$, which is the inverse frequency of peak $S(f)$, and $-1$ moment period, also known as energy mean wave period, $T_{0m1}=\frac{\int_{f0}^{f1} f^{-1}S(f)df}{m_0}$). 
The notations are frequency spectrum $S$  and frequency $f$. The integration range was ($f0$,$f1$)  for $n$th spectral moments $m_n$ where $f1=0.308$ Hz.
Zeni-v2021 data were affected by the noise floor that was elevated while the buoy was drifting in open water until around 25 September; the elevated noise floor is analogous to the \citet{Waseda2017,Waseda2018,Nose2018} observation and seems to accompany IMU-based wave sensors housed in a relatively small floating platform. The same ideal filter method as \citet{Waseda2017,Waseda2018,Nose2018} was implemented to derive the wave statistics. As such, $f0$ was not a constant and depended on the ideal-filter cut-off frequency. The Zeni-v2021 integration range ($f0$,$f1$) were matched in the SPOT-1386 wave statistics calculation. It is noteworthy that the elevated noise floor was observed until 25 September, which roughly coincides with the time when the inertial oscillation visible in the Inset A of Fig. \ref{fig1} seemingly stopped (indicated by the white crosses in Figs. \ref{fig1} and \ref{fig2}).  

\textbf{Inter-buoy comparison:}
It is apparent in the $H_{m0}$ panel of Fig. \ref{fig2} that, despite the buoys having similar drifting trajectories for the first half of the deployment, $H_{m0}$ began to vary slightly between them after half a day and considerably after just two days. As we will show throughout the paper, the variability likely indicates that the wave evolution was modified by the sea ice fields via one of the following effects: 1). waves are attenuated as they propagate into the ice cover medium. 2). lateral boundary conditions are imposed by the ice fields and affects the wave evolution over the effective fetch.  

Although we discuss the possibility that the effective fetch at the buoys' location after 12 hours of deployment was already affected by the sea ice lateral boundary in Section~\ref{lateralboundary}, we aim to consolidate the general inter-buoy agreement when the buoys were in close proximity. Scatter for $H_{m0}$ and $T_{0m1}$ are plotted in Fig. \ref{fig3}. The markers were grouped by an arbitrary buoy distance threshold of 5 km to demonstrate that the buoys' wave statistics agreed better when the distance between them was shorter, i.e., the effect of the sea ice field on wave evolution is 	less for shorter distances. Indeed, the blue markers, indicating the data when the buoy distance was < 5 km, in both panels are clustered closer to the black dotted agreement line than the red markers. Furthermore, as was shown in the left panel of \citet{Rabault2022}'s Fig. 7, the spectra agreed well immediately after the buoys were deployed.

While precise Zeni-v2021 validation with SPOT-1386 were impeded by the growing sea ice fields, we showed that the Zeni-v2021 measurement quality seems sufficiently adequate when the wave evolution was less altered by the sea ice field (when the distances between them were short). As such, the wave events observed by the co-located buoy measurements were used to evaluate the operational ARC MFC wave-ice model predictability in Section \ref{iceeffectsonwaves}.

\section{Numerical models}\label{model}
\subsection{ARC MFC wave-ice model}\label{ARCMFCmodel}
\textbf{The wave-ice interaction parameterisation}: The ARC MFC wave-ice model is an operational wave model product for the Arctic Ocean and includes a wave-ice interaction parameterisation, i.e., the model can simulate wave propagation in sea ice cover. Medium-range forecast and analysis are distributed via the CMEMS platform.

The ARC MFC wave-ice interaction is based on \citet{Sutherland2019}; 
they modelled wave dissipation with a 2-layer sea ice model: the top layer is modelled like a thin film that has no horizontal motion with thickness (1-$\epsilon) h_i$ while the bottom layer is modelled as a moving viscous layer. Here, $h_i$ is the ice thickness. \citet{Sutherland2019} describe that the $\epsilon$ coefficient is related to ice permeability at the microscopic scale and is a function of ice temperature, salinity, and ice volume fraction. 
The assumption of a highly viscous top layer is similar to \citet{Weber1987}, who derived a well-known wave dissipation solution by modelling the thin ice cover as an inextensible layer that halts the horizontal motion of the fluid layer underneath. 
The \citet{Sutherland2019} model could be considered an extension of the \citet{Weber1987} model. In the \citet{Weber1987} model, the dissipation rate is $\alpha \propto K_{35}f^{3.5}$ where $K_{35}$ is a constant. The \citet{Sutherland2019} model also has a frequency dependence, but they also included an ice thickness dependence to the dissipation rate like $\alpha \propto K_{40}f^4$ where $K_{40}$ is a function of ice thickness. A more thorough discussion on the various dissipation rates in the literature is provided in \citet{Waseda2022}.

The \citet{Sutherland2019} model was developed based on a scaling argument as they derived that the viscosity scales with SIT, $\nu \propto h_i$, which led to spatial wave dissipation as a function of ice thickness and frequency: $\alpha \propto h_i f^{4}$. The dissipation rate is parameterised in Eq. 16 of \citet{Sutherland2019} as 
\begin{equation}\label{alpha}
	\alpha = \frac{1}{2}\Delta_0 \epsilon h_i k^2  
\end{equation}
where $\Delta_0 \approx 1$.
In the CMEMS Quality Information Document\textsuperscript{+},\footnote{\textsuperscript{+}\url{https://catalogue.marine.copernicus.eu/documents/QUID/CMEMS-ARC-QUID-002-014.pdf} accessed 30 Nov 2022.} the dissipation rate is denoted $\alpha = C_d h_i k^2$ where they state that $C_d$ is a tuning parameter and determined by the best fit to observation obtained from the ice covered fjord, Tempelfjorden, at Svalbard in 2018.
Note that when $\epsilon=0$ (or $C_d$), the model may behave like the Weber model \citep{Weber1987} if the boundary layer in the fluid is implemented (not done so in \cite{Sutherland2019} yet).
Lastly, the dissipation rate $\alpha$ is used to estimate the dissipated wave spectrum like
\begin{equation}\label{eq:dissipation}
	S(f) = S_0(f)e^{-\alpha x}
\end{equation} 
where $S_0$ is the incoming wave spectrum and $x$ is a distance between the two points.

\textbf{The wave part} of the ARC MFC wave-ice model is based on Met Norway's version of WAM \citep{WAMDI1988}.  WAM is a spectral wave model that is discretised in frequency and direction, and solves the numerical evolution of ocean waves as energy budgets based on the action density balance equation. The surface wind boundary conditions are forced using the ECMWF HRES atmospheric forecast. 
At the ocean boundary along 53$^\circ$ N latitude, wave lateral boundary conditions are directional wave spectra from the ECMWF HRES wave forecast, also based on WAM. 
The ECMWF HRES atmospheric forecast has a regular lon/lat grid at 0.1 deg. 
The ARC MFC wave-ice model has a spatial resolution of 3 km on the polar stereographic projection. 

\textbf{The sea ice part} of the ARC MFC wave-ice model is taken from the ARC MFC ocean analysis\textsuperscript{*}\footnote{\textsuperscript{*}CMEMS product name: ARCTIC\_ANALYSIS\_FORECAST\_PHYS\_002\_001\_A}. 
The sea ice model is the CICE and based on the viscous–plastic sea ice rheology. The sea ice model uses a 1-thickness category sea ice model based on the thermodynamics described in \cite{Drange1996} \citep{Sakov2012}. There, the minimum thickness for the newly formed ice is given as 0.5 m. 
Implications of the minimum thickness value is evaluated with the buoy observations in Section \ref{discordantthickness}.

The CICE sea ice model is coupled to the HYCOM. The atmospheric forcing is obtained from the ECMWF HRES atmospheric forecast. The data assimilation is performed weekly using the ensemble Kalman filter \citep{Sakov2012} and for the following: altimeter sea level, in situ temperature and salinity profiles, the Operational Sea Surface Temperature and Ice Analysis sea surface temperature, OSI-SAF sea ice concentration and drift observations, and CryoSAT2 and Soil Moisture and Ocean Salinity ice thickness (in winter).  The HYCOM has a horizontal resolution of \textasciitilde12 km, which is more than 3 times the resolution of the ARC MFC wave-ice model.

\subsection{ECMWF HRES wave forecast}\label{ECMWFmodel}
Wave heights and periods from the ECMWF HRES wave forecast were used in this study as a tool to analyse the ARC MFC wave-ice model and the buoy observation comparison. 
The ECMWF HRES wave forecast was obtained for our research activity by the Arctic Data archive System (\url{https://ads.nipr.ac.jp/}). We took a series of 0--24 hour forecasts to produce a time series during the observation period. The ECMWF wave model accounts for sea ice using ice masks where grid cells with SIC > 0.30 are treated as land.
The ECMWF HRES wave forecast model has a regular lon/lat grid at 0.125 deg.
The ECMWF wave model is useful to evaluate the wave conditions in the open water near the ice edge based on the following logic:
\begin{enumerate}
	\item When the wave evolution occurs entirely over the open water fetch,  the ECMWF wave and ARC MFC wave-ice models should agree.
	\item Near the ice edge, where satellite-derived sea ice data are uncertain, inaccurate sea ice representation can cause erroneous wave predictions \citep{Nose2020}.
	\item In such cases, the ECMWF wave model that neglects sea ice may produce better predictions.
\end{enumerate}
%As will be shown in Section \ref{lateralboundary}, the ECMWF wave model serves as a useful tool to evaluate wave conditions near the ice edge.

\subsection{neXtSIM sea ice model} \label{neXtSIMmodel}
The neXtSIM sea ice model\textsuperscript{+}\footnote{\textsuperscript{+}CMEMS product name: ARCTIC\_ANALYSISFORECAST\_PHY\_ICE\_002\_011. Data downloaded on 19 Jan 2022.} 
\citep{Rampal2016,Olason2022} is a sea ice product distributed by the CMEMS and based on the Brittle-Bingham-Maxwell rheology.  
%\cite{Williams2021} describe that the neXtSIM model was developed to improve the dynamical aspects, like the ice drift and deformation statistics and scaling, that classical sea ice models based on viscous–plastic and elastic-viscous–plastic rheologies perform poorly.  
\cite{Rampal2016} introduced this model describing that fracturing and faulting of sea ice should be expressed as an assembly of plates > O(1 km) and floes O(100 m), rather than an intact solid plate. 
%Furthermore, because the new ice formed in leads can fuse with broken ice and contribute to the effective mechanical recovery of ice, they also adapted the thermodynamics from that of the classical models. 
The neXtSIM thermodynamical component is based on a 3-category model that includes open water, newly formed ice, and older ice as described in \cite{Rampal2019}. 
%The adapted thermodynamics may be a relevant feature to our study; 
\cite{Rampal2016} shows their model calculates the ice formation in the newly formed ice category based on the atmosphere and ocean forcing, whereas a prescribed growth rate is adopted conventionally in classical models \citep{Rampal2016}. The thickness range of this newly formed ice category is from 0.05 m to 0.275 m, which is considerably thinner than that of the ARC MFC ocean analysis.

Unlike the ARC MFC ocean model, neXtSIM is not coupled to an ocean circulation model, but is coupled with a mixed-layer model that is relaxed to an ocean circulation model.
The ocean part is represented by a single level "slab ocean" model of the mixed layer \citep{Rampal2016} and uses the following daily averaged forcing from the TOPAZ4 ocean data assimilation model system: sea surface (0--3 m) ocean velocity, temperature and salinity, and the mixed layer depth \citep{Williams2021}.
The atmospheric forcing is the ECMWF HRES atmospheric forecast, and the neXtSIM model assimilates OSI-SAF SIC via the nudging scheme on a daily basis. Since the control variable is only the SIC, neXtSIM is strongly constrained to the observation, i.e., the OSI-SAF SIC.
The model has a Lagrangian triangular mesh with an equivalent square grid resolution of \textasciitilde7 km, for which the data are distributed on the 3 km polar stereographic projection grid.

\section{The effects of sea ice on wave evolution as observed by the wave buoys and models}\label{iceeffectsonwaves}
\subsection{Sea ice as lateral boundary effects on wave evolution}\label{lateralboundary}
Significant wave height $H_{m0}$ and period time series were extracted from the ARC MFC wave-ice model and the ECMWF HRES forecast at the Zeni-v2021 and SPOT-1386 positions during their deployment between 15 and 29 September 2021. The full time series for wave heights and periods are shown in Figs. \ref{figA2} and \ref{figA3} of \ref{appA}.  In this subsection, we focus on the wave heights between 15 and 19 September immediately after the buoys were deployed in open water (see Fig. \ref{fig4}). 

Following the deployment, Fig. \ref{fig4} shows that the ECMWF $H_{m0}$ agrees well with both buoys for up to \textasciitilde12 hours as denoted by the magenta bars, whereas the ARC MFC $H_{m0}$ is clearly underestimated. Shortly after, however, the buoys' $H_{m0}$ takes a steep decrease at around 14:00 15 September, as indicated by the green arrows in Fig. \ref{fig4}, and the model-observation trend reversed, i.e., the ECMWF $H_{m0}$ is overestimated and the ARC MFC $H_{m0}$ agree qualitatively with the observation. It is also noted that this change coincides with the time when the Zeni-v2021 $H_{m0}$ began deviating slightly from that of SPOT-1386 (blue dotted line in Fig. \ref{fig2}) as discussed in Section \ref{colocatedbuoy}.

The ARC MFC wave fields immediately after the deployment at 06:00 15 September are shown in Panel (a) of Fig. \ref{fig5}. The grey vectors are the mean wave directions scaled by $T_{0m1}$ values. At this time, it can be seen that wave directions were primarily parallel to the ice edge. Tracing upwind from the buoys, the ice edge protrude and shelters the buoys' effective fetch on which waves can grow. Considering the ARC MFC $H_{m0}$ is underestimated, the ice edge sheltering of the buoys may be the cause of the underestimation, i.e., the protruding ice edge may not be an accurate representation of the sea ice field. For reference, Panel (c) of Fig. \ref{figA4} in \ref{appB} shows the buoys are not sheltered by any protruding ice edge features in the ECMWF sea ice representation. 

Panel (b) of Fig. \ref{fig5} shows the ARC MFC wave fields at 00:00 16 September after the models' trend reversed, i.e., now the ARC MFC $H_{m0}$ has better agreement than the ECMWF $H_{m0}$. The fetch orientation transitioned from along the ice edge to off-ice wave conditions. The Panels (b) and (d) of Fig. \ref{figA4} in \ref{appB} show that winds were also blowing from the ice cover. As indicated by the purple bars in Fig. \ref{fig5}, the ECMWF $H_{m0}$ is clearly overestimated and the ARC MFC $H_{m0}$ agrees with the observations with a varying degree of predictability. 

The observations here demonstrate the lateral boundary effect that sea ice imposes on the wave fields. We conjecture that inaccurate representation of the ARC MFC ice tongue imposed a lateral boundary condition that prohibited reproducing the true wave evolution when the fetch was orientated parallel to the ice edge. A similar scenario was also observed on 21 September 2021, in which the ice tongue representations between the ARC MFC and ECMWF models were markedly different. This event is described in \ref{appC} to present further support to this conjecture. 

\subsection{Wave dissipation due to sea ice on 29 September}
South to southeasterly winds over the Laptev Sea generated on-ice waves on 29 September 2021. The model wave fields are shown in Fig. \ref{fig6}. By this time, both Zeni-v2021 and SPOT-1386 were in dense ice cover with SIC > 0.8 (see Fig. \ref{fig2}). The ice edge geometry and wave orientation were not straightforward;  the Zeni-v2021 location was downwind compared to the SPOT-1386, but the Zeni-v2021 distance to the ice edge relative to the wind direction was closer than that of SPOT-1386. This is depicted in the wave field shown in Fig.~\ref{fig6}. 

During this event, the ECMWF wave model has no values at both buoys as they are covered by the ice mask. The ARC MFC wave-ice model simulates no waves at both buoys (see Fig.~\ref{figA2} of \ref{appA}). 
However, the $H_{m0}$ time series panel in Fig. \ref{fig2} depicts that waves were observed at Zeni-v2021 reaching a peak $H_{m0}$ value over 1.25 m. SPOT-1386 located \textasciitilde30 km southeast of Zeni-v2021 did not observe any waves. 
To investigate how the wave energy propagated to the downwind Zeni-v2021, but not for SPOT-1386, we analysed the model representations of the sea ice field of the ARC MFC ocean analysis and the neXtSIM sea ice model product. 

\section{Sea ice model representation differences and their effects on wave evolution} \label{discussion}
\subsection{Comparison of ARC MFC and neXtSIM sea ice field representations} \label{icefieldcomp}
Sea ice fields between ARC MFC and neXtSIM are compared at the buoys' location on 29 September 2021 in Fig. \ref{fig7}. It is apparent that the horizontal scale of ARC MFC sea ice features (shown in Panels (a) and (c)) are effectively limited to the ARC MFC ocean analysis scale of \textasciitilde12 km. 
This may be expected as interpolation to a high resolution grid is unlikely to reveal features less than the original scale; the implication here is that there is an inconsistent scale between the wave-ice model geographical configuration and the model physics resolution. The scale of sea ice features in the neXtSIM sea ice model appears to be much finer; the reason for this may be speculated that the Lagrangian nature of the model can influence the thermodynamics as well as the mechanical properties.

Notwithstanding this, the striking differences between the two models pertaining to the buoys' wave observation are the ice edge location, the 0.80 SIC contour, and the sea ice thickness differences. 
The neXtSIM SIC fields shows that the distance from Zeni-v2021 to the sea ice edge is much closer than shown in the ARC MFC field. Moreover, if we take the wave propagation as roughly 150 degrees, then, 0.80 SIC ice-covered sea that the waves need to propagate to reach Zeni-v2021 is much shorter than that of the ARC MFC wave-ice model, while for SPOT-1386, this distance remains relatively similar.
Regarding the sea ice edge and 0.80 SIC contour differences between the ARC MFC ocean analysis and the neXtSIM sea ice model, we revisit the data assimilation intervals and scheme differences: the ARC MFC ocean analysis assimilates data at weekly intervals and uses the ensemble Kalman filter with many control variables (see Section~\ref{ARCMFCmodel}) whereas the neXtSIM sea ice model carries out daily data assimilation via the nudging scheme with SIC being the only control variable (see Section~\ref{neXtSIMmodel}). The data here suggest that the difference in the data assimilation intervals and schemes produce diverging sea ice field representations.

The representation of SIT in the ARC MFC model for thin ice appears to be poor: the ice field beyond the 0.80 SIC contour is practically SIT > 0.3 m. By contrast, the neXtSIM SIT field clearly has a thickness distribution between 0 m and 0.3 m within the plot domain.
\ref{icechart} describes the observation-based ice chart that confirms the ice type near the ice edge and the wave buoys was young ice. 
Although it is beyond the scope of this paper to examine whether coupling a wave model with the neXtSIM sea ice forcing reproduces the co-located buoy data, there is sufficient evidence to suggest that the neXtSIM sea ice representation appears to be in accord for reproducing the Zeni-v2021 $H_{m0}$, i.e., less dissipation because of low SICs and thinner thickness.

\subsection{Disparate scale between wave dissipation parameterisation and the sea ice thickness forcing}\label{discordantthickness}
%We claimed that the poor representation of the ARC MFC SIT less than 0.3 m is an adverse limitation in the preceding section, which requires corroboration. 
The SIT disparity between the two models at the regional scale is also shown in Fig. \ref{figA6} of \ref{appD}, which indicate the ARC MFC SIT representation up to 0.5 m is poor.  
Sea ice thickness is one of the most fundamental sea ice variables, yet it remains difficult to measure as reliable methods are via ice core sampling, Electromagnetic–induction instruments, and select satellite observations \citep{Tateyama2006,Tilling2018}.
In other words, regional/synoptic scale estimation of a SIT field is typically not readily available.
For wave-ice models, an implication is that ice thickness can often serve as a wave dissipation tuning parameter. For ocean-ice coupled models, the essential feedback of sea ice to the ocean is thermodynamics, which is rather insensitive to thin ice. As such, one of the tuning parameters may be the minimum thickness parameter; indeed, the ARC MFC thermodynamic model is based on \cite{Drange1996}, which has the minimum thickness of newly formed ice as 0.5 m.

Another possibility for the poor thin ice representation is the data assimilation method. The AFC MFC ocean analysis is based on the ensemble Kalman filter scheme \citep{Sakov2012} and does not assimilate sea ice thickness in the summer months. As shown in the previous section, the ice field beyond the 0.80 SIC contour is practically SIT > 0.3 m in Fig. \ref{fig7}; it is plausible that ARC MFC ocean analysis data assimilation may assume correlation between the observed SIC and unobserved SIT, which could be the cause of the poor thin ice representation.

From the wave-ice interaction viewpoint, however, pancake ice thickness is typically 10 cm, and new and young ice thickness is less than 30 cm (\url{http://old.aari.ru/resources/nomen/volume1.php?lang1=0&lang2=1&arrange=0&self=0}). The ARC MFC wave-ice interaction parameterisation is based on the \citet{Sutherland2019} two layer model, which parameterised the wave dissipation rate to scale with the sea ice thickness. Accordingly, the poor representation of MIZ thin ice thickness in the sea ice forcing is a critical issue.  
Based on Fig. \ref{fig7}, and Fig. \ref{figA6} of \ref{appB}, it appears that thin thickness distribution that is appropriate for the \cite{Sutherland2019} wave dissipation model is not resolved in the ARC MFC ocean analysis.

According to \citet{Sutherland2019} Eq 16, $\epsilon=1$ determines the maximum dissipation in the moving viscous ice layer like $\alpha = \frac{1}{2} h_i k^2$. The maximum wave dissipation rates along the approximate wave propagation were plotted in Fig. \ref{fig8} for the wave period of 7 s to quantify the effects of SIT differences between the ARC MFC and neXtSIM models. The SIT was extracted for 200 km along the approximate wave propagation ray from 121$^\circ$E, 82.2$^\circ$N on an initial bearing of 150 degrees. For relevance, $T_{p}=$7 s was roughly the upper limit of the co-located buoy observation when wave energy was detected (see Fig. \ref{fig2}).
For SIT between the transect distance 25--75 km,  Fig. \ref{fig8} shows that the ARC MFC wave dissipation rate can exceed 3 times than that of the neXtSIM counterpart. This figure confirms that the ARC MFC ocean analysis ice thickness is not adequately resolving the thin ice to apply the dissipation model of \cite{Sutherland2019}; this prohibited a meaningful model-observation comparison to evaluate the \citet{Sutherland2019} model with our wave buoy observation, e.g., can their SIT dependent dissipation rate reproduce waves in thin ice-covered MIZs.
For completeness, we conducted an academic experiment to show how $\frac{d H_{m0}}{dx}$ differ due to poor thin SIT resolution using a Pierson Moskowitz spectrum with a 7.5 s $T_p$ that has a 2.2 m $H_{m0}$. The dissipated $H_{m0}$ was calculated from the spectrum $S$ using Eq. \ref{eq:dissipation}. 
Assuming a maximum dissipation rate over a 10 km distance with thickness 0.1 m and 0.5 m, we estimate $H_{m0} =0.1$ and $0.6$ m, respectively. The difference of $\frac{d H_{m0}}{dx}$ for the 0.5 m thickness results in 2.6 times the dissipation than the 0.1 m thickness case. However, it is difficult to know how this is reflected in the ARC MFC wave-ice model as the true SIC field, the sea ice edge, and the incoming spectra shape are all unknown.

Based on the discussion here, it is clear the ARC MFC wave-ice model during our buoy observation overestimated the wave dissipation in the thin ice fields with SIT less than 0.5 m. The ARC MFC wave-ice model in MIZs near the ice edge needs sea ice thickness forcing that can resolve the thin ice types.
It is worthy to mention that the neXtSIM sea ice model has been experimentally coupled with the wave model of \cite{Boutin2021}, however, with a different viewpoint: their focus was the effect of waves on 	ice fragmentation utilising the brittle rheology of neXtSIM.

\section{Conclusions}\label{conclusion}
Two drifting wave buoys were deployed in the central Arctic Ocean, north of the Laptev Sea, where there have not been any wave observations available.
The motivation for the buoy deployments was to validate Zeni-v2021, a prototype of an experimental wave-ice buoy named OMB \citep{Rabault2022}, with SPOT-1386, a commercial wave buoy. As such, the buoys were deployed at the same location in the open water adjacent to an ice edge at a time ocean surface freezes and new ice forms. 
We quickly learnt that as the buoy positions deviated, so did their wave heights, which prevented a full validation of Zeni-v2021. However, the inter-buoy comparison showed that when the buoy distances between Zeni-v2021 and SPOT-1386 were close, defined here with the arbitrary threshold of 5 km, Zeni-v2021 was sufficiently accurate compared to the commercial SPOT-1386 buoy. As such, the buoy data were used to study the predictability of waves in the deployment region.

The first event we focused was shortly after the deployment when a sudden decrease in the buoys' significant wave heights $H_{m0}$ from \textasciitilde1.75 m to \textasciitilde1.50 m was observed. The decrease coincided with the change in wind directions from along the ice edge to off-ice wind. We compared the observation with an operational ARC MFC wave-ice model product and found that the model underestimated $H_{m0}$ before the sudden decrease, but agreed better afterwards. Tracing upwind of the fetch, we found that the ice tongue constrained the wave growth over the available fetch. We conjecture the ice tongue location was inaccurate because the buoy observation agrees well with the ECMWF HRES wave forecast before the sudden decrease, and there is no ice tongue upwind of the fetch in the ECMWF sea ice forcing. This shows that the representation of the ice edge is critically important for accurate predictions of waves in the nearby open water.

The second case we analysed was when the wave buoys entered ice cover as new ice formed in the area. An on-ice wave event generated waves that propagated to the buoys when they were \textasciitilde30 km apart. Zeni-v2021 was located downwind of SPOT-1386, but was slightly closer to the ice edge relative to the wind direction. During this event, Zeni-v2021 measured waves with a peak $H_{m0}$ of 1.25 m, but SPOT-1386 did not detect any waves. To investigate how the wave energy propagated to Zeni-v2021 but not for SPOT-1386, we compared the buoy observation with the ARC MFC wave-ice model.
Because the ARC MFC model simulates waves in ice with a wave dissipation model developed for thin ice covered seas \citep{Sutherland2019}, we expected reasonable model agreement; however, the ARC MFC wave-ice model $H_{m0}$ was largely underestimated. With a viewpoint to elucidate the model error, we examined the model sea ice conditions  between the ARC MFC ocean analysis and the neXtSIM sea ice model product.
The analysis showed that the ARC MFC sea ice forcing does not resolve thin thickness distribution for ice types like new and young ice with typical thickness less than 30 cm. Since the ARC MFC wave-ice model's dissipation rate has a SIT dependence, the ARC MFC model overestimates wave dissipation in the thin thickness ice field. On the other hand, we found that another CMEMS sea ice product, neXtSIM, seemed to better resolve the thin thickness distributions in the MIZ.
One of the reason for the difference maybe that the neXtSIM sea ice model calculates the new ice formation using the atmosphere and ocean forcing \citep{Rampal2019} rather than using the conventional approach of adopting a prescribed growth rate with a minimum thickness for new ice is configured as 0.5 m \citep{Drange1996}.
Another factor is the data assimilation; while the ARC MFC data assimilation is sophisticated with many control variables, it is plausible that its data assimilation may be assuming correlation between the observed SIC and unobserved SIT (when SIT is not observed in the summer months) as the ice covered areas above 0.80 SIC seems to have 0.5 m SIT or higher. 
Regardless of the causes, the ARC MFC wave-ice model needs sea ice forcing that reproduces thin ice thickness cover for better predictions of ocean waves in MIZs when new ice is forming.

Reliable ocean wave forecasts are crucial for safe navigation in the polar seas. Our study presented an observational insight into the wave and ice model coupling for better predictions of waves in MIZ thin ice cover.
Observations and models help us ensure sustainable developments, such as safe navigation, of the changing Arctic Ocean.

\nolinenumbers
\singlespacing
\section*{Acknowledgements}
We sincerely thank 2021 NABOS chief scientist Dr Igor Polyakov for supporting our wave buoy deployment.\\
We are grateful to Dr Guillaume Boutin and the anonymous reviewer for providing constructive criticisms during the peer review. Their invaluable comments led to a much improved manuscript.\\
The authors are grateful to CMEMS for providing a data distribution platform. We are also sincerely thankful to the model developers of the ARC MFC wave-ice model and the neXtSIM sea ice model.\\

\section*{Funding}
Drs N. Kanna and K. Tateyama were supported by the ArCS II Japan-Russia-Canada International Exchange Program to participate in the 2021 NABOS expedition.\\
This work was a part of the Arctic Challenge for Sustainability II (ArCS II) Project (Program Grant Number JPMXD1420318865). A part of this study was also conducted under JSPS KAKENHI Grant Numbers JP 19H00801, 19H05512, 21K14357, and 22H00241.\\
%------------------------------------------------

%----------------------------------------------------------------------------------------
\clearpage
\section{Figures}
%\import {../} {figure_vrev}
\begin{figure}[h]
	\centering
	\includegraphics[width=\textwidth]{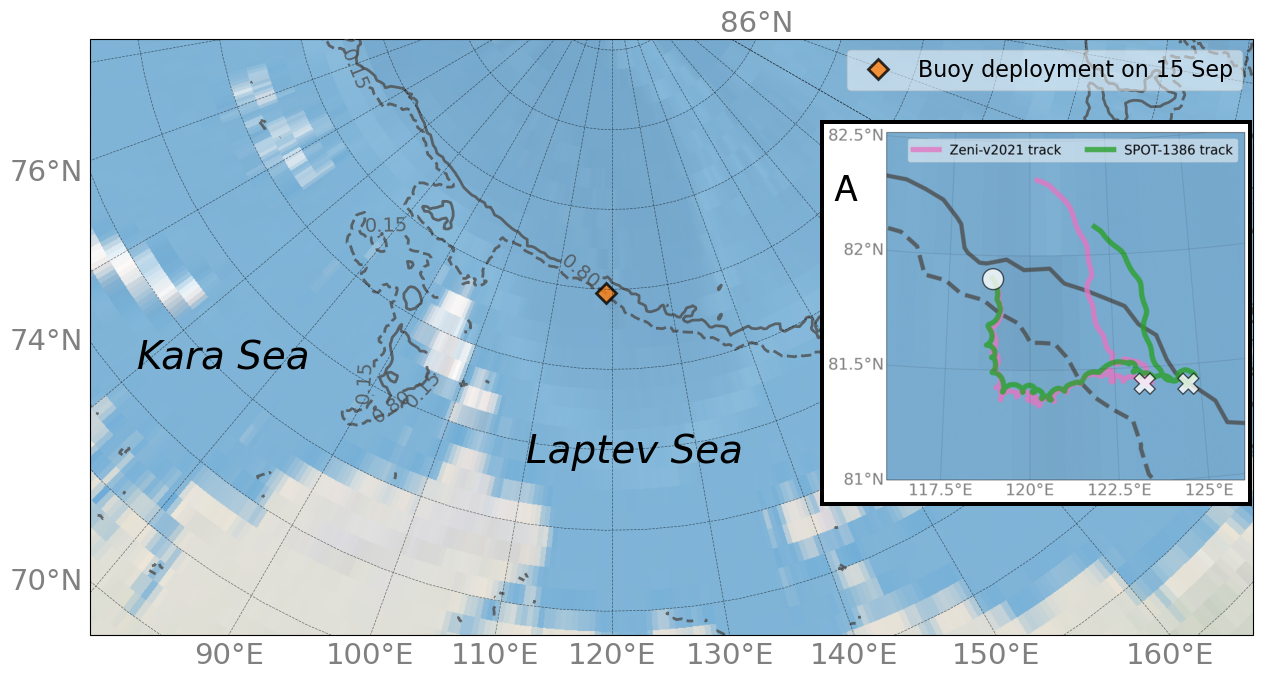}
	\caption{A location map of the Arctic Ocean where the co-located wave buoys were deployed on 15/09/2021. The dashed and solid lines indicate the 0.15 and 0.80 AMSR2 SIC contours on the same day. Inset A shows the Zeni-v2021 (pink) and SPOT-1386 (green) trajectories between 15 and 29 September, in which the white circle indicate the deployment location. The white crosses show approximate location when the visible inertial oscillation stops on 25/09/2021, which could indicate the change in ocean surface conditions. The 0.15 and 0.80 AMSR2 SIC contours for 25/09/2021 are overlaid in Inset A.}
	\label{fig1}
\end{figure}

\begin{figure}[h]
	\centering
	\includegraphics[width=\textwidth]{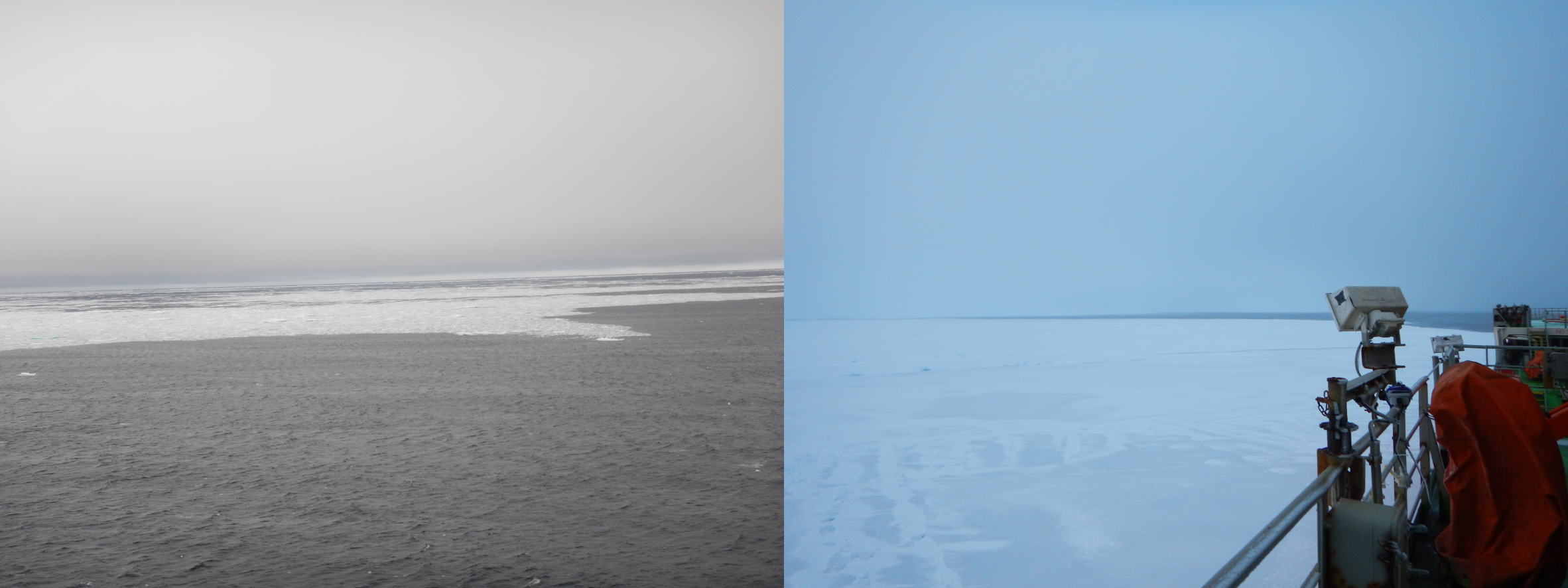}
	\caption{Sea ice conditions encountered by R/V Akademik Tryoshnikov on 15/09/2021 after the buoys were deployed. The left image is showing the grey ice while the right image is showing the grey-white ice; both ice types belong to the young ice category.}
	\label{figA1}
\end{figure}

\begin{figure}[h]
	\centering
	\includegraphics[width=\textwidth]{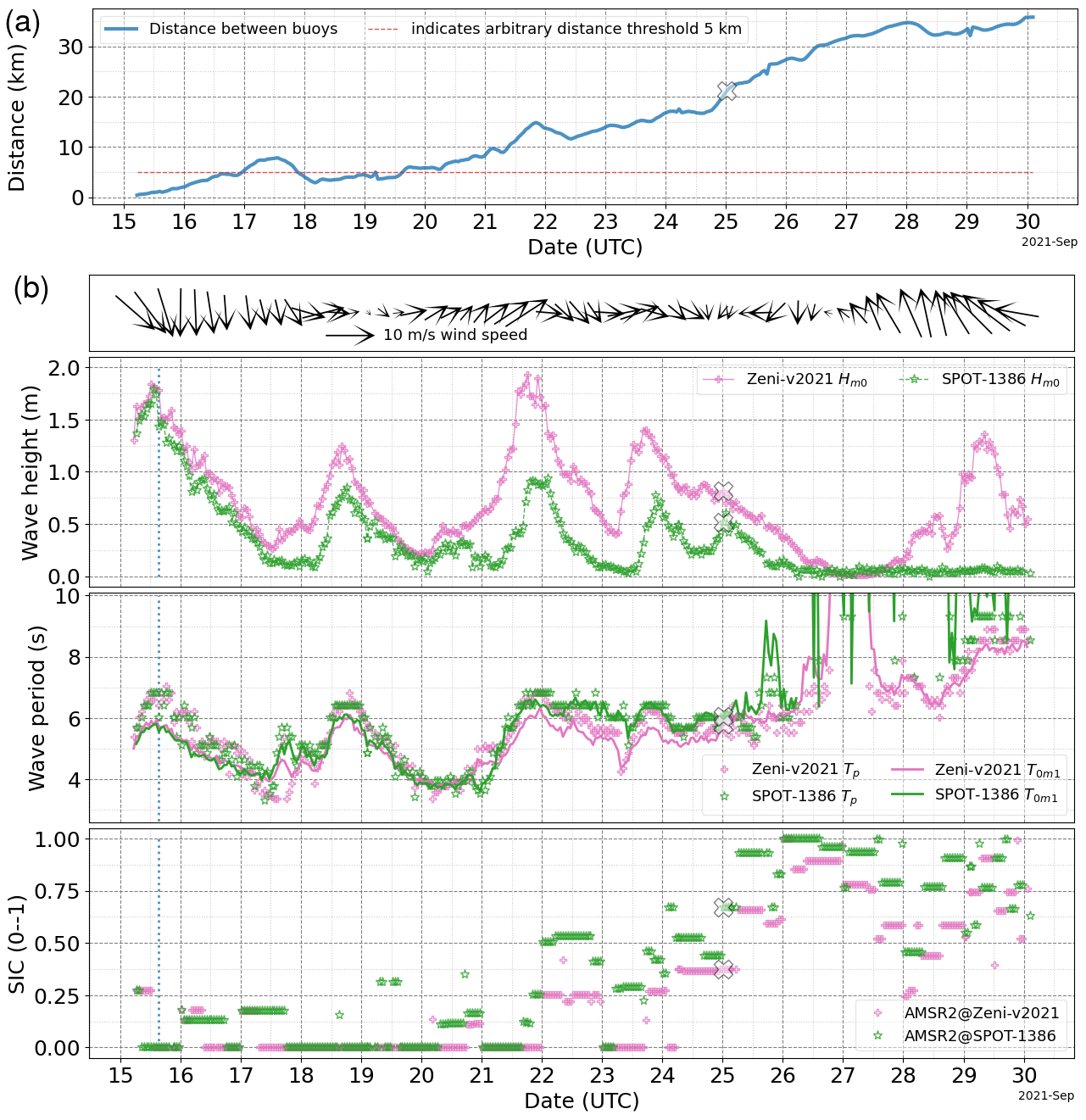}
	\caption{An overview of the co-located wave buoy measurements. Panel (a) is a time series of buoy distances. In Panel (b), the top panel shows the ECMWF HRES atmosphere forecast wind vector extracted at the Zeni-v2021 location. The middle panels show the buoy significant wave height $H_{m0}$ (upper) and energy mean and peak wave periods $T_{0m1}$, $T_p$ (lower). The bottom panel shows the AMSR2 SIC extracted at the buoy locations. The blue dotted lines indicate when the buoys' $H_{m0}$ began to deviate slightly (discussed in Sections \ref{colocatedbuoy} and \ref{lateralboundary}). The white crosses in this figure corresponds to the approximate times when the inertial oscillations seemingly stopped as shown in Fig. \ref{fig1}.
	}
	\label{fig2}
\end{figure}

\begin{figure}[h]
	\centering
	\includegraphics[width=\textwidth]{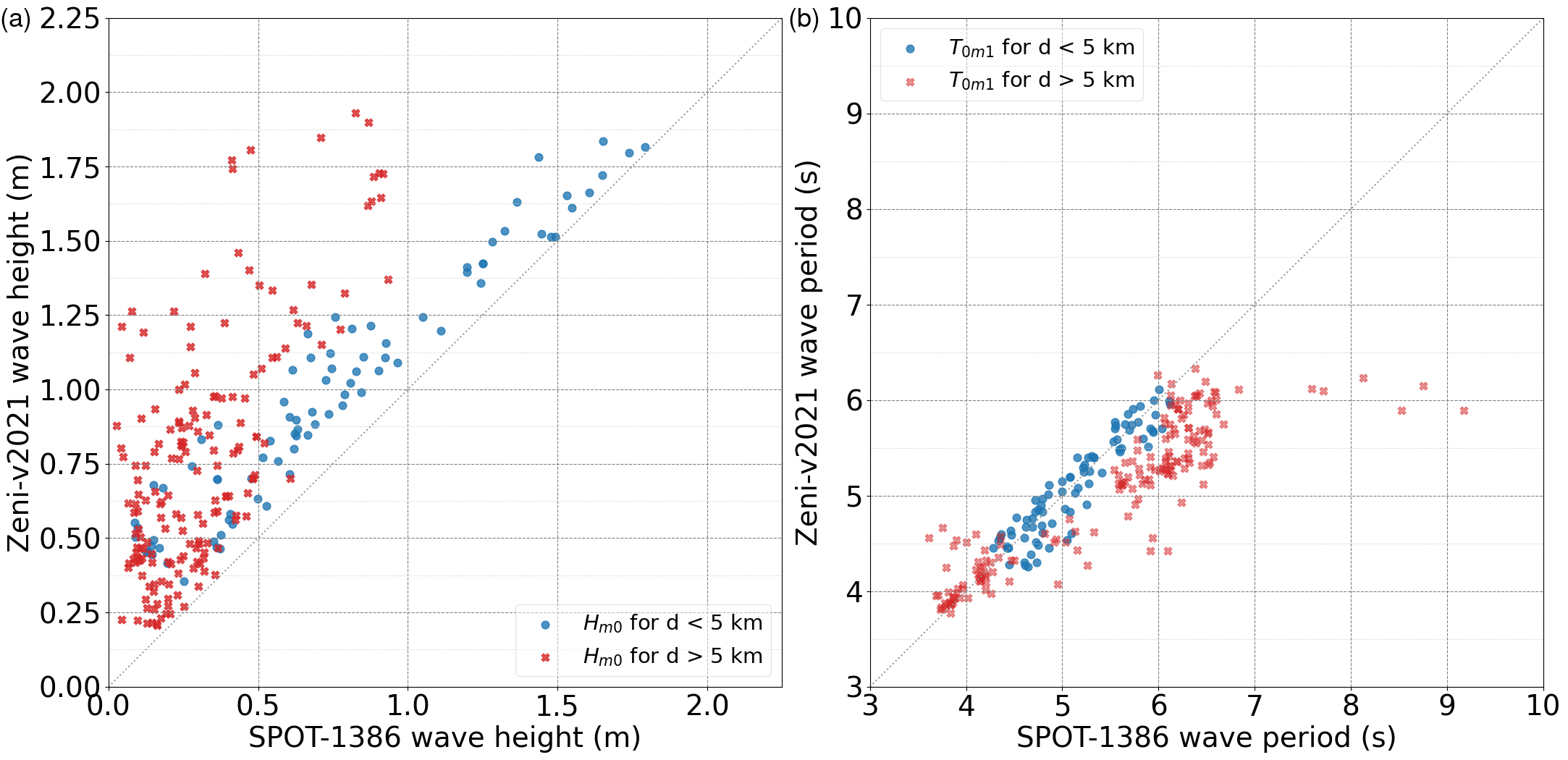}
	\caption{Scatter plots of Zeni-v2021 and SPOT-1386 from 15 to 25 September. The latter date was chosen to avoid comparing wave periods that were affected by the buoy noise floor (see the wave periods shown in Fig. \ref{fig2}). Panel (a) shows the $H_{m0}$ (m) and Panel (b) shows the $T_{m01}$~(s). The marker colours were grouped using an arbitrary buoy distance threshold where blue and red are less and greater than 5 km apart, respectively. The black dotted lines are the agreement line, and blue markers tend to cluster to these lines. }
	\label{fig3}
\end{figure}

\begin{figure}[h]
	\centering
	\includegraphics[width=\textwidth]{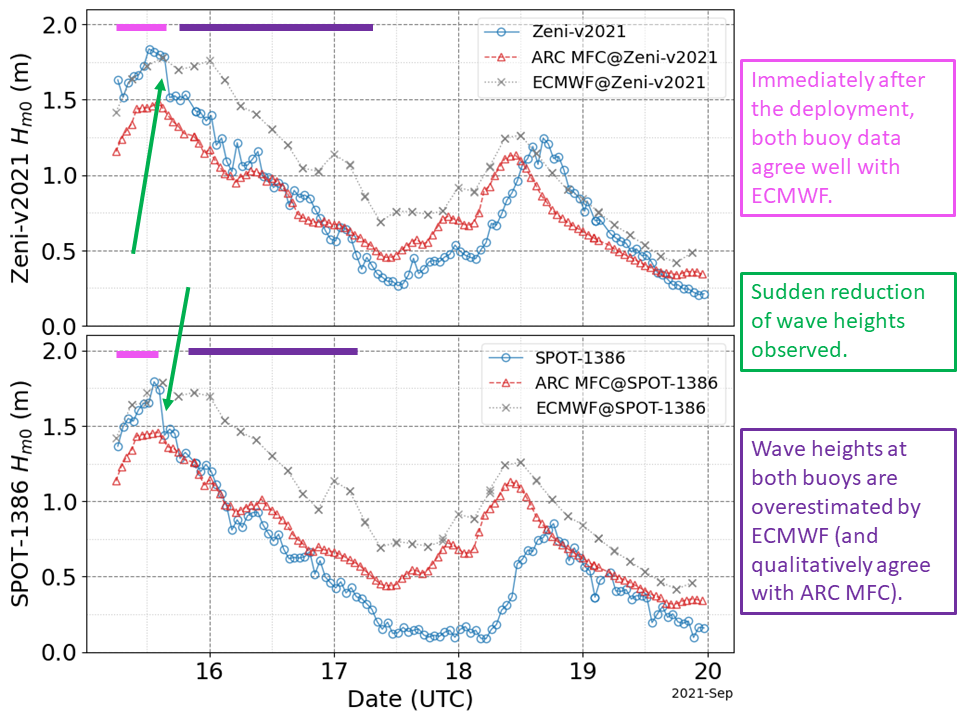}
	\caption{Significant wave height $H_{m0}$ time series comparing the buoy observation (blue) and the models between 15 and 19 September 2021 when the buoy distances were generally less than 5 km after the deployment. Both the ARC MFC wave-ice model (red) and the ECMWF HRES wave forecast (grey) are shown. The missing values in the ECMWF HRES wave forecast is due to ice masks (grid cells with SIC > 0.30). The top panel compares the Zeni-v2021 data while the bottom panel compares SPOT-1386. The magenta bars and text indicate when the buoy observation agreed with the ECMWF HRES wave forecast. The green arrows show when a sudden reduction in the wave heights were observed. The purple bars and text indicate when the buoy observation agreed qualitatively with the ARC MFC wave-ice model.}
	\label{fig4}
\end{figure}

\begin{figure}[h]
	\centering
	\includegraphics[width=\textwidth]{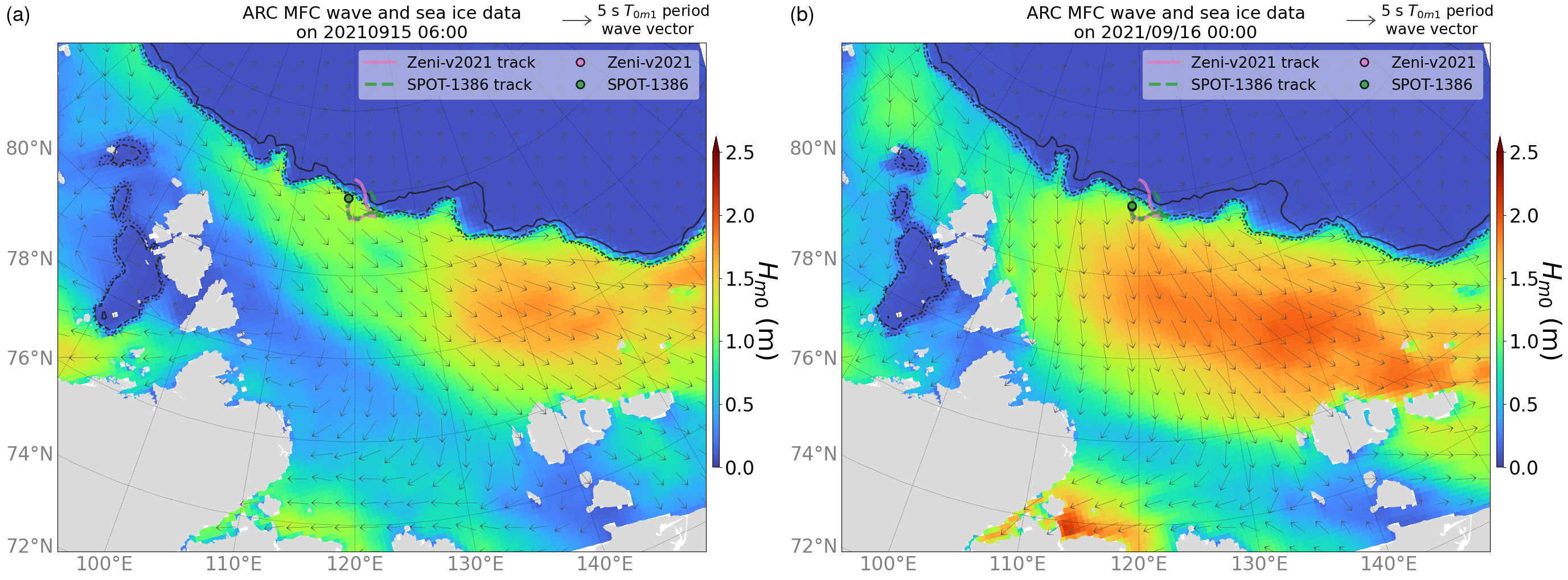}
	\caption{ARC MFC wave fields showing how the fetch orientation changed from along the ice edge immediately after the deployment at 06:00 15/09/2021 as shown in Panel (a) to off-ice by 00:00 16/09/2021 as shown in Panel (b). Both figures show the $H_{m0}$ via colour while the grey vectors correspond to mean wave directions with the vector lengths scaled by $T_{0m1}$. Both panels show the Zeni-v2021 (pink) and SPOT-1386 (green) trajectories and positions, and are overlaid with the 0.15 (dotted), 0.30 (dashed), and 0.80 (solid) SIC contour lines.}
	\label{fig5}
\end{figure}

\begin{figure}[h]
	\centering
	\includegraphics[width=\textwidth]{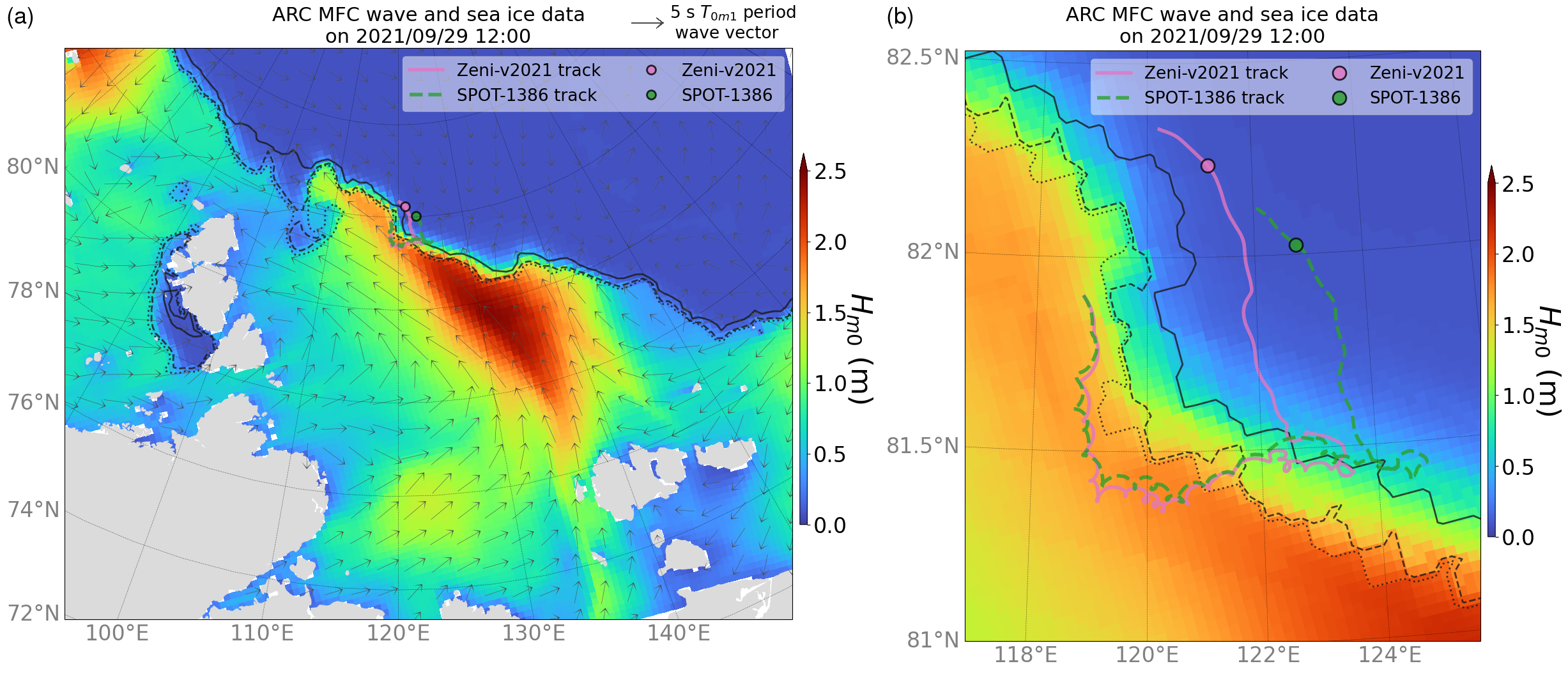}
	\caption{ARC MFC ice and wave fields during the wave event on 29/09/2021 when Zeni-v2021 measured $\sim$1.3 m $H_{m0}$, but the model showed no waves. Panel (a) indicates the wave conditions in which the colours indicate $H_{m0}$ while the grey vectors correspond to mean wave directions with the vector lengths scaled by $T_{0m1}$. Panel (b) is a zoomed view of Panel (a) near the buoys. Both figures show the Zeni-v2021 (pink) and SPOT-1386 (green) trajectories and positions, and are overlaid with the 0.15 (dotted), 0.30 (dashed), and 0.80 (solid) SIC contour lines.}
	\label{fig6}
\end{figure}

\begin{figure}[h]
	\centering
	\includegraphics[width=\textwidth]{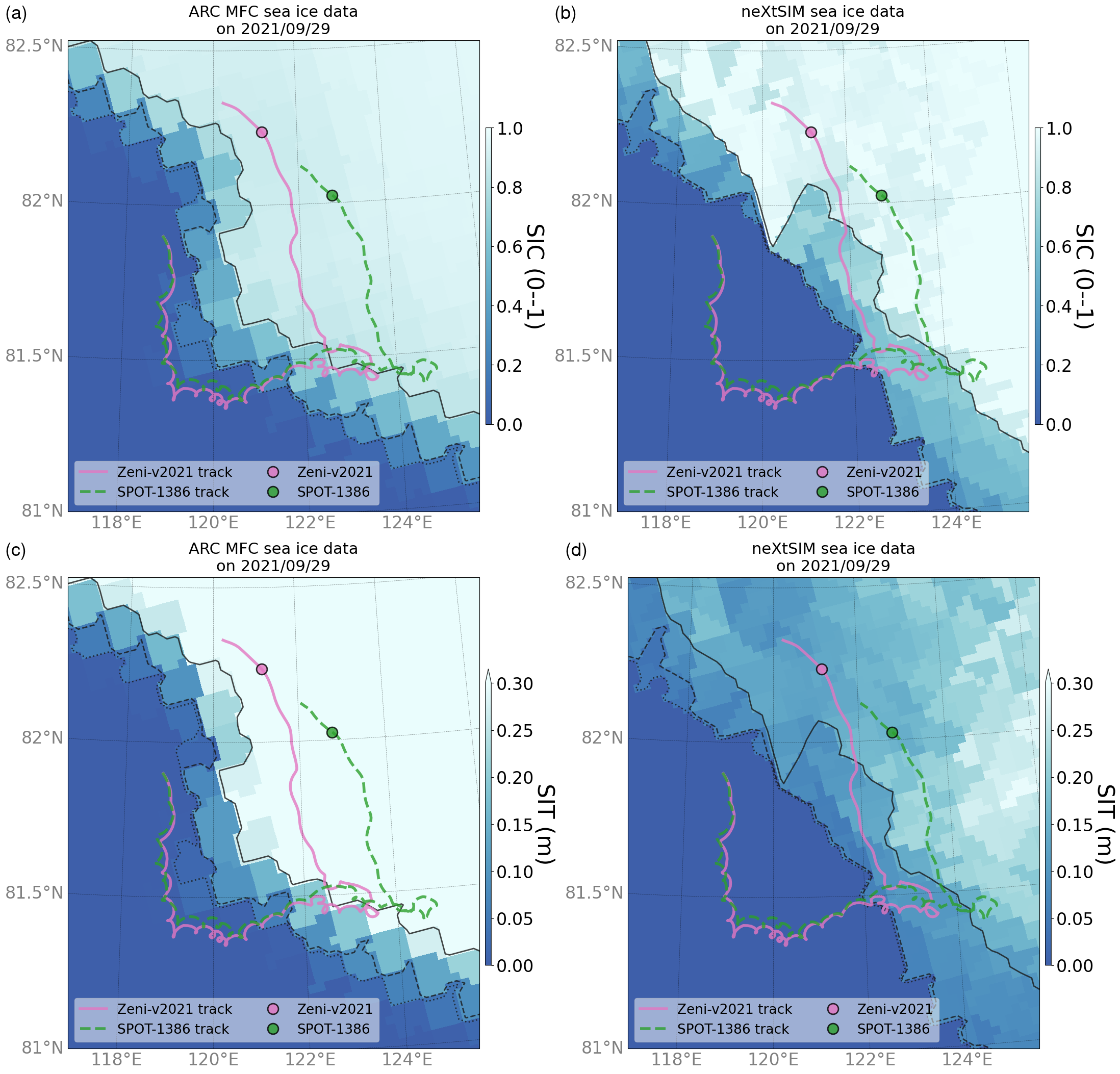}
	\caption{Comparison of ice conditions on 29/09/2021 between the ARC MFC (left figures in Panels (a) and (c)) and neXtSIM models (right figures in Panels (b) and (d)). The top figures (Panels (a) and (b)) show the SIC and the bottom figures (Panels (c) and (d)) show the SIT. All figures show the Zeni-v2021 (pink) and SPOT-1386 (green) trajectories and positions, and are overlaid with the respective 0.15 (dotted), 0.30 (dashed), and 0.80 (solid) SIC contour lines.}
	\label{fig7}
\end{figure}

\begin{figure}[h]
	\centering
	\includegraphics[width=\textwidth]{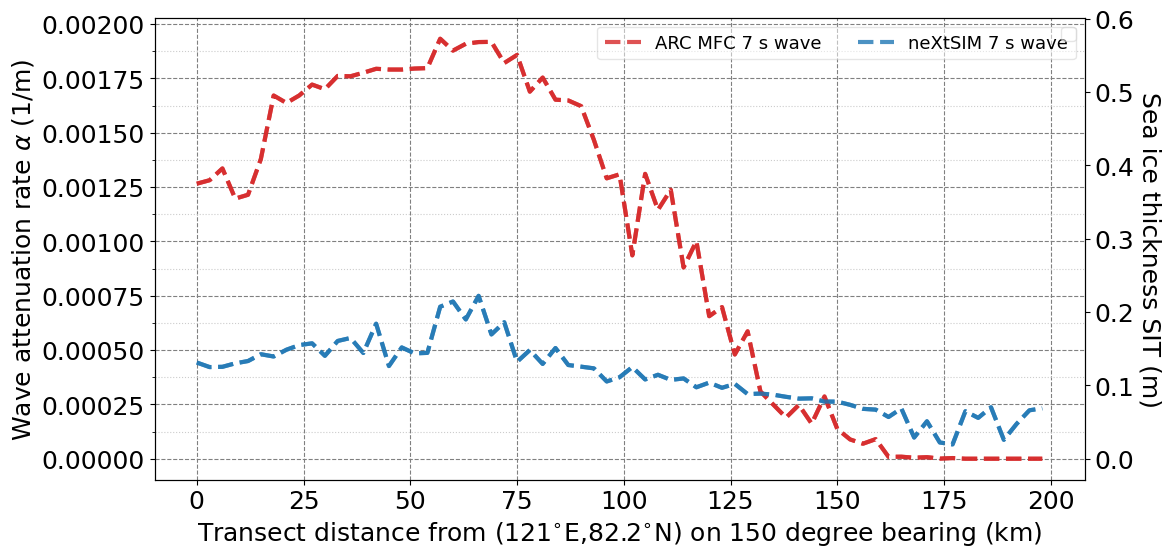}
	\caption{Comparison of \citet{Sutherland2019} wave dissipation rate $\alpha = \frac{1}{2} \epsilon h_i k^2$ where $\epsilon=1$ based on the ARC MFC (red) and neXtSIM (blue) model sea ice thickness $h_i$. $\alpha$ is plotted along a 200 km transect from 121$^\circ$E, 82.2$^\circ$N at an initial bearing of 150 degrees for waves with 7~s periods. The right axis is the sea ice thickness.}
	\label{fig8}
\end{figure}
\clearpage
%----------------------------------------------------------------------------------------
%	BIBLIOGRAPHY
%----------------------------------------------------------------------------------------

\renewcommand{\refname}{\spacedlowsmallcaps{References}} % For modifying the bibliography heading

\bibliographystyle{abbrvnat}
\bibliography{./references} % The file containing the bibliography

\clearpage
\appendix
%\import {../} {appendix_vrev}
\setcounter{figure}{0}
\renewcommand{\thefigure}{A\arabic{figure}}
\renewcommand*{\thesection}{Appendix~\Alph{section}}
\section{Appendix}
\subsection{Ice type near the ice edge} \label{icechart}
Arctic and Antarctic Research Institute (AARI) ice charts during the 2021 NABOS expedition were available on \url{http://wdc.aari.ru/datasets/d0040/arctic/png/2021/}. The regional ice charts are updated monthly and available on \url{https://aari.ru/data/realtime} (although only available in the Russian language). The pan-Arctic and regional ice charts were obtained to estimate the ice type near the buoys during the 29 September event, and these are shown in Fig. \ref{figA7}. The buoys were located around $\sim$82$^\circ$ N,	 $\sim$122$^\circ$ E during the event, and it appears to show that the ice type near the ice edge is the young ice type.
Note that the pan-Arctic ice chart was obtained from \url{http://wdc.aari.ru/datasets/d0040/arctic/png/2021/blended_arcice_20210930-20211005_sd_90E.png} and the regional ice chart from \url{http://old.aari.ru/odata/_d0004.php?mod=0&m=Lap} (by selecting year 2021 and month/day 2021.10.05 on the right side boxes).

\begin{figure}[h]
	\centering
	\includegraphics[width=\textwidth]{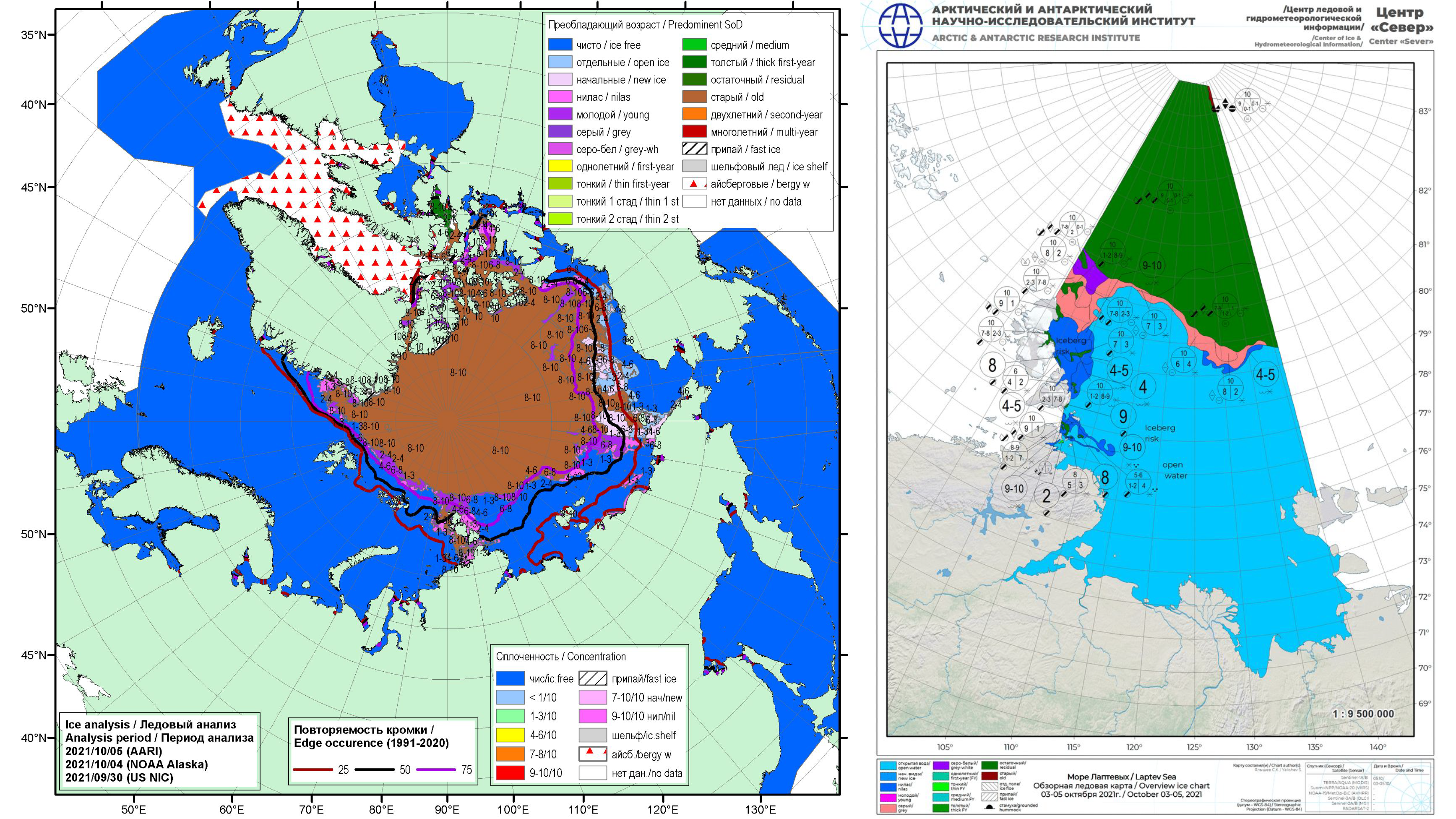}
	\caption{AARI ice charts for the stage of development for the pan-Arctic (left panel) and regional (right panel) waters around the 29 September event are shown. The buoys were located around $\sim$82$^\circ$ N, $\sim$122$^\circ$ E during the 29 September event, and it shows that the ice type near the ice edge was young  ice.}
	\label{figA7}
\end{figure}    
\clearpage

\subsection{Time series comparison of observed and modelled waves}\label{appA}
Significant wave height $H_{m0}$ and wave periods $T_p$ and $T_{0m1}$ time series were extracted from the ARC MFC wave-ice model and the ECMWF HRES wave forecast at the Zeni-v2021 and SPOT-1386 positions during their co-located deployment between 15 and 29 September 2021. The time series are plotted in Figs. \ref{figA2} and \ref{figA3}, respectively, and supports the Section \ref{iceeffectsonwaves} text. The ECMWF HRES wave forecast adopts ice masks, which treat grid cells with SIC > 0.30 as land. 
From the five $H_{m0}$ peaks captured, as shown in Fig. \ref{figA2}, the models were not able to reproduce reasonable values at both buoys simultaneously, and they could only reproduce $H_{m0}$ to a varying degree of accuracy only at one of the buoys.

\begin{figure}[h]
	\centering
	\includegraphics[width=\textwidth]{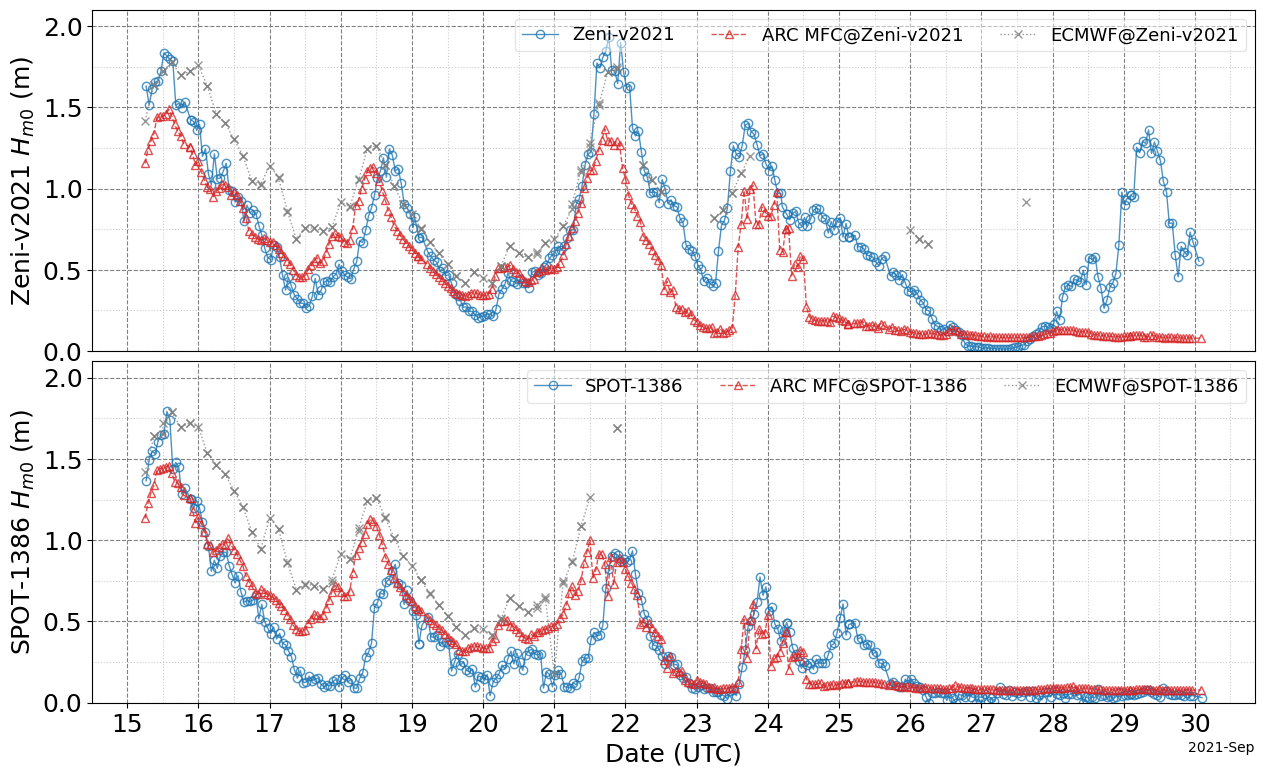}
	\caption{Significant wave height $H_{m0}$ time series comparing the buoy observation (blue) and the models during the co-located measurements between 15 and 29 September 2021. Both the ARC MFC wave-ice model (red) and the ECMWF HRES wave forecast (grey) are shown. The missing values in the ECMWF HRES wave forecast is due to ice masks (grid cells with SIC > 0.30). The top panel compares the Zeni-v2021 data while the bottom panel compares SPOT-1386.}
	\label{figA2}
\end{figure}

\begin{figure}[h]
	\centering
	\includegraphics[width=\textwidth]{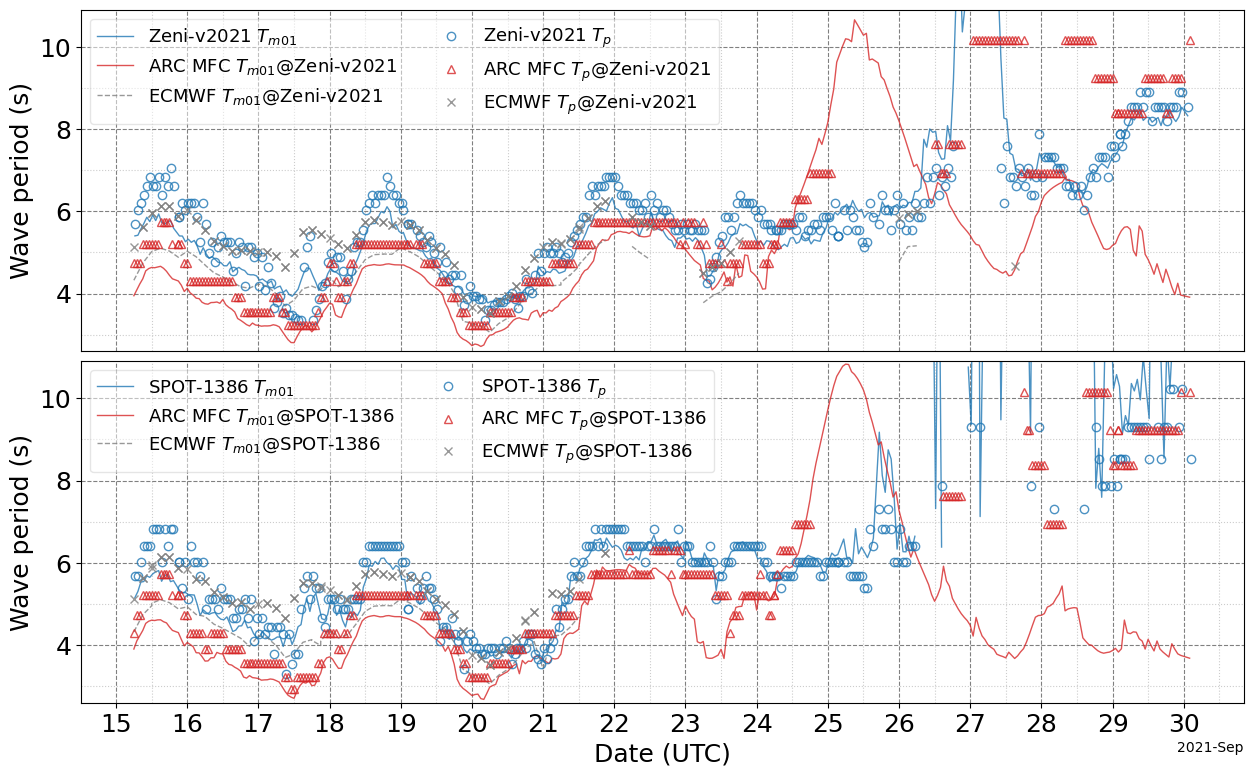}
	\caption{Figure \ref{figA2} equivalent for wave periods comparing the buoy observation (blue) with the models.  $T_p$ is shown as markers and $T_{0m1}$ shown as lines. Both the ARC MFC wave-ice model (red) and the ECMWF HRES wave forecast (grey) are shown. The missing values in the ECMWF HRES wave forecast is due to ice masks (for grid cells with SIC > 0.30). The top panel compares the Zeni-v2021 data while the bottom panel compares SPOT-1386.}
	\label{figA3}
\end{figure}

\clearpage
\subsection{Lateral boundary effects of sea ice}\label{appB}
The wind and sea ice fields for the 15--19 September event are shown in Fig. \ref{figA4} to support the Section \ref{lateralboundary} text. Comparing the left and right panels, it can be seen that the fetch orientation changed from along the ice edge (left panels) to off-ice (right panels). As discussed in Section \ref{lateralboundary}, the ARC MFC representation of the sea ice field has a protruding ice edge that sheltered the wave buoys from wave evolution along the ice edge immediately after the buoy deployments. By contract, the ECWMF ice edge representation is smooth, and the wave evolution towards the buoy appears to not be sheltered by any ice edge feature. 

\begin{figure}[h]
	\centering
	\includegraphics[width=\textwidth]{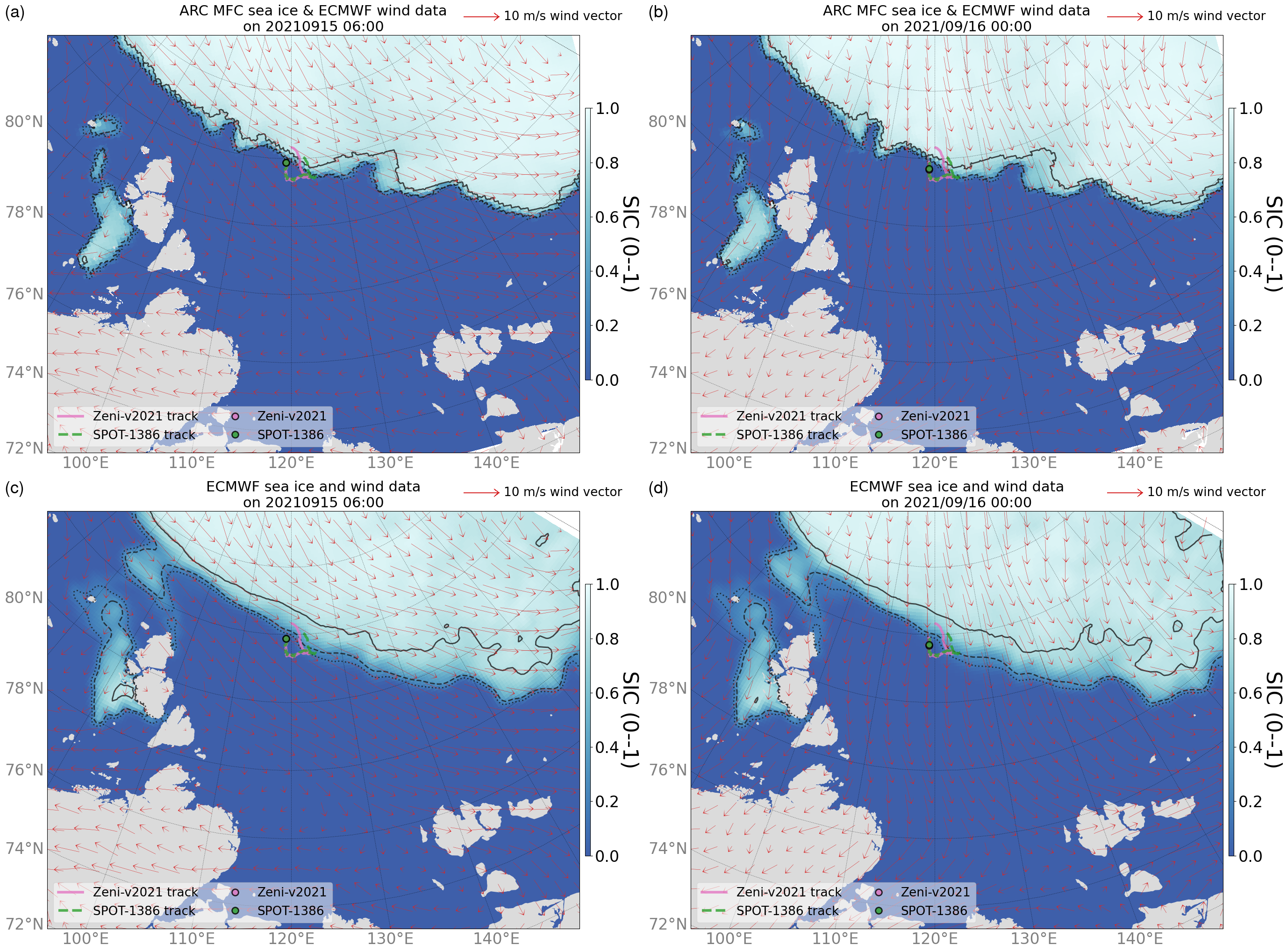}
	\caption{The figure shows how the fetch orientation changed from along the ice edge immediately after the deployment at 06:00 15/09/2021 (Panel (a) and (c)) to off-ice by 00:00 16/09/2021 (Panel (b) and (d)). Both figures show the SIC via colour while the red vectors correspond to the ECMWF wind field. The top panels show the ARC MFC sea ice conditions while the bottom panels show the ECMWF sea ice conditions.
		All panels show the Zeni-v2021 (pink) and SPOT-1386 (green) trajectories and positions, and are overlaid with the respective 0.15 (dotted), 0.30 (dashed), and 0.80 (solid) SIC contour lines.}
	\label{figA4}
\end{figure}

\subsection{Lateral boundary conditions sheltered wave evolution at the Zeni-v2021 possibly due to misrepresentation of an ice tongue}\label{appC}
On 21 September, west to southwest winds generated on-ice waves, i.e., waves propagating towards the ice edge, that peaked with a $H_{m0}$ value of almost 2 m at Zeni-v2021. At this time, SPOT-1386 was located closer to the ice edge than Zeni-v2021, and its $H_{m0}$ only peaked at $\sim$1 m. 

It can be seen in Fig. \ref{figA2} that Zeni-v2021 $H_{m0}$ agrees reasonably with the ECMWF HRES wave forecast whereas the ARC MFC wave-ice model somehow underestimates the $H_{m0}$. A snapshot of wind, ice, and wave conditions for the ECMWF and ARC MFC models are provided in Fig.~\ref{figA5}. In the ARC MFC wave field, the Zeni-v2021 position is seaward of the 0.1 SIC contour, so the underestimation is not caused by anomalous attenuation due to ice. 
Rather, the ECMWF and ARC MFC SIC fields as shown in the top panels in Fig. \ref{figA5} depict the location of the ice tongue are inconsistent. The ARC MFC SIC field shows the ice tongue was located near the 110$^\circ$ E. Based on the wind and SIC fields, it can be conjectured that the different representation of the ice tongue affected the open water fetch; in this particular case, the ARC MFC ice tongue location was likely inaccurate considering the ECMWF $H_{m0}$ agreement with that of Zeni-v2021. 

\begin{figure}[h]
	\centering
	\includegraphics[width=\textwidth]{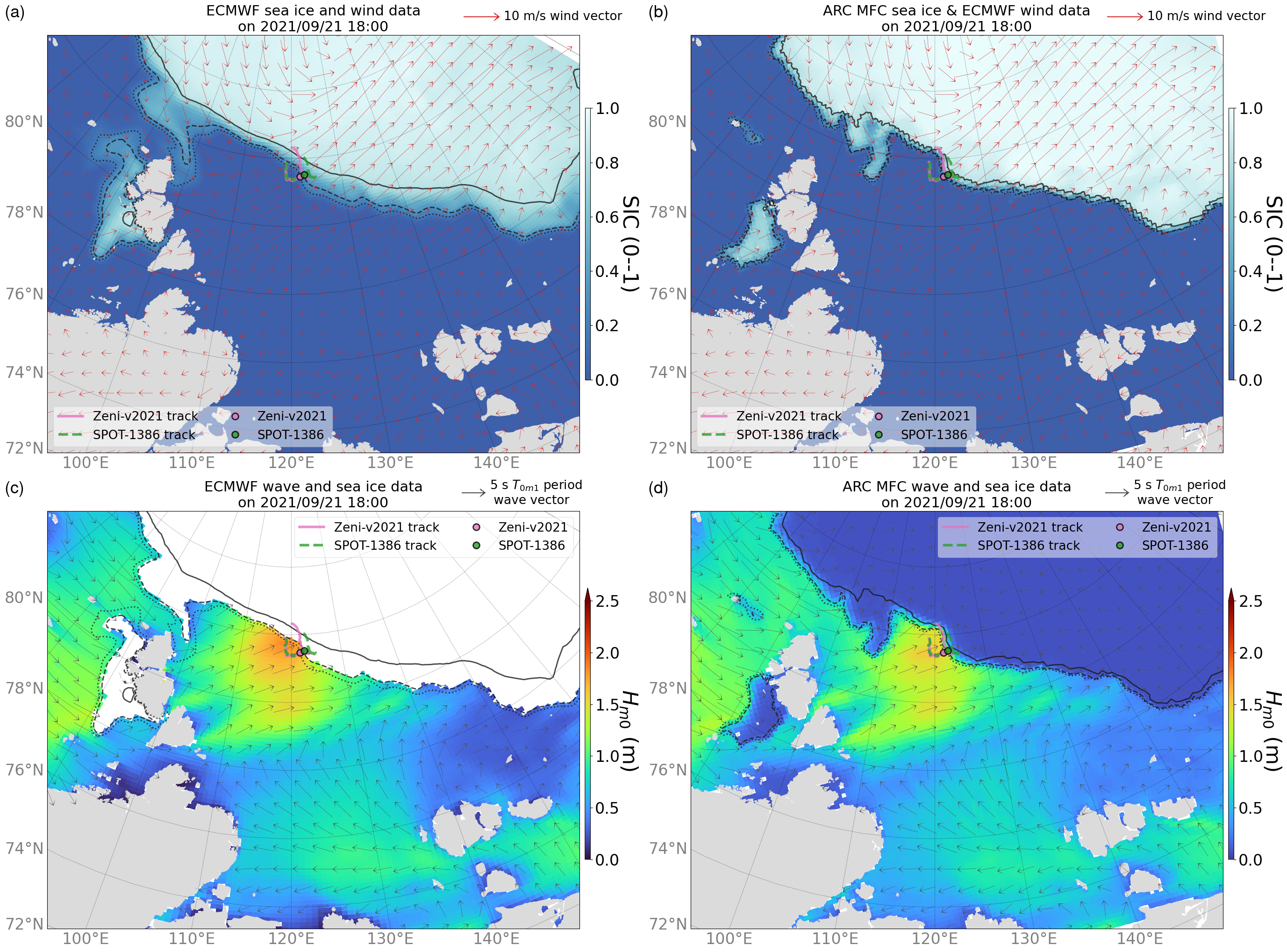}
	\caption{The figure shows the different representation of the ice tongue between the ECMWF and ARC MFC models on 18:00 21/09/2021. The top panels (Panels (a) and (b)) shows the wind (red vectors) and ice (colours) conditions and the bottom panels (Panels (c) and (d)) shows the waves. The left panels (Panels (a) and (c)) correspond to the ECMWF data while the right panels show the ARC MFC data, although the wind data are the ECMWF HRES atmospheric forecast for both top panels. All panels show the Zeni-v2021 (pink) and SPOT-1386 (green) trajectories and positions, and are overlaid with the respective 0.15 (dotted), 0.30 (dashed), and 0.80 (solid) SIC contour lines.
		In the bottom wave figures, the colours correspond to $H_{m0}$ while the grey vectors indicate the mean wave directions, for which vector lengths are scaled by the corresponding $T_{0m1}$. }
	\label{figA5}
\end{figure}

\subsection{Disparate sea ice thickness distributions between ARC MFC and neXtSIM sea ice fields} \label{appD}
A comparison of ARC MFC and neXtSIM sea ice fields during the 29 September at the regional scale is presented in Fig. \ref{figA6} to support the text in Section \ref{discordantthickness}. It is apparent that ARC MFC sea ice thickness resolution below 0.5 m is considerably poorer than that of neXtSIM. It is conjectured in the main text that this may be due to the following. neXtSIM consist of a newly formed ice category in which the ice formation is calculated from the atmosphere and ocean forcing, whereas the ARC MFC model uses a 1-thickness category model in which the minimum thickness of newly formed ice is set as 0.5 m \citep{Drange1996,Sakov2012}. We also discuss another possible contributing factor, which is the data assimilation method. The poor thin ice distribution may be caused by the data assimilation method assuming correlation between observed SIC and unobserved SIT in the ARC MFC ocean analysis.

\begin{figure}[h]
	\centering
	\includegraphics[width=\textwidth]{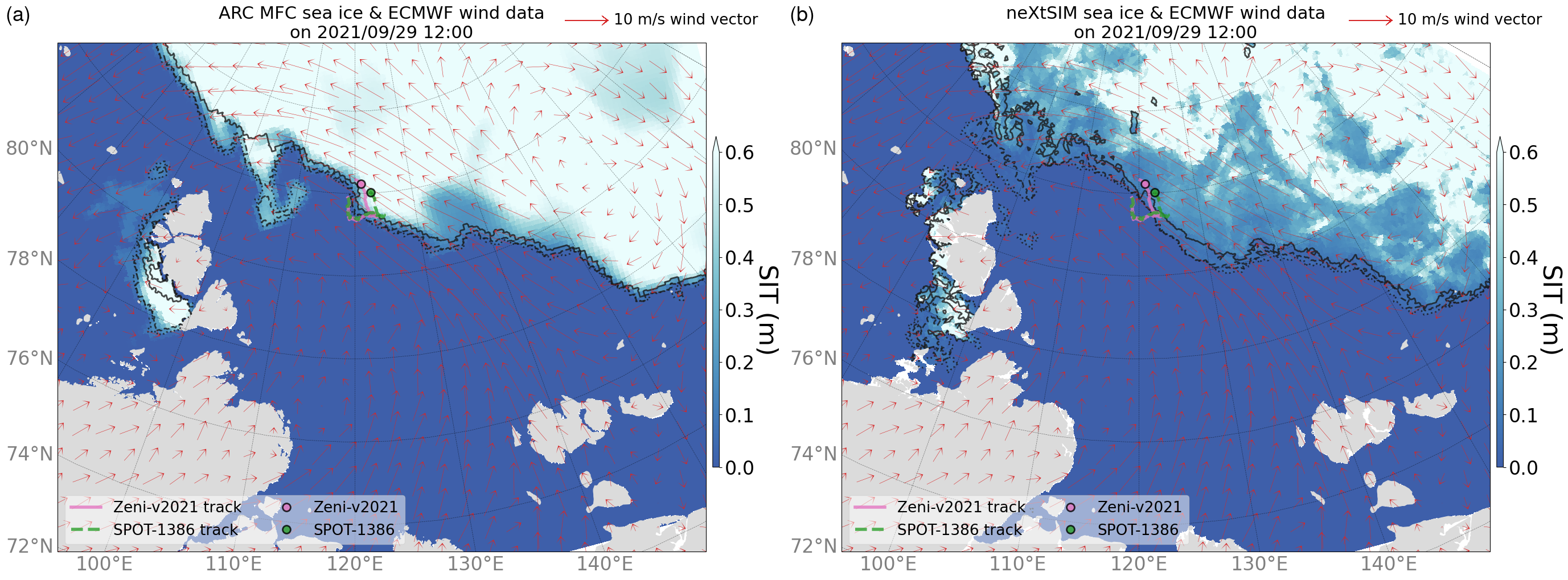}
	\caption{A comparison of sea ice thickness (colours) fields for the ARC MFC wave-ice model sea ice forcing (Panel (a)) and the neXtSIM sea ice model (Panel (b)) at a regional scale. The figure shows that there is a marked difference in the thin ice thickness distributions between the two models. The red vectors are the ECMWF model wind data. Both figures show the Zeni-v2021 (pink) and SPOT-1386 (green) trajectories and positions, and are overlaid with the respective 0.15 (dotted), 0.30 (dashed), and 0.80 (solid) SIC contour lines. }
	\label{figA6}
\end{figure}

\end{NoHyper}
\end{document}